\documentclass[prd,superscriptaddress,amsfonts,amssymb,amsmath,showpacs,twocolumn,nofootinbib]{revtex4-2}
\usepackage{bm}
\usepackage{amsfonts}
\usepackage{latexsym}
\usepackage{graphicx}
\usepackage{amsmath}
\usepackage{palatino}
\usepackage{xcolor} 
\usepackage{mathpazo}
\usepackage{tensor}
\usepackage{textcomp}
\usepackage{float}
\usepackage{booktabs}
\usepackage{dcolumn}
\usepackage{booktabs}
\usepackage{multirow}
\usepackage{hyperref}
\hypersetup{colorlinks,citecolor=blue}
\usepackage{amsmath}
\usepackage{xcolor}
\usepackage{orcidlink}
\usepackage{epsfig}
\usepackage{caption}
\usepackage{subcaption}
\usepackage{commath}
\hypersetup{colorlinks,citecolor=blue}
\hypersetup{colorlinks=true,linkcolor=magenta,filecolor=magenta,    urlcolor=blue}

\def\be{\begin{equation}}
\def\ee{\end{equation}}
\def\bea{\begin{eqnarray}}
\def\eea{\end{eqnarray}}

\begin{document}

\title{\bf $f\left(R,\mathcal{G},T\right)$ gravity: cosmological implications, and a dynamical system analysis}

\author{Ratul Mandal}
\email{ratulmandal2022@gmail.com} 
\affiliation{Department of
Mathematics, Indian Institute of Engineering Science and
Technology, Shibpur, Howrah-711 103, India}
\author{Himanshu Chaudhary}
\email{himanshu.chaudhary@ubbcluj.ro,\\
himanshuch1729@gmail.com}
\affiliation{Department of Physics, Babeș-Bolyai University, Kogălniceanu Street, Cluj-Napoca, 400084, Romania}
\author{Tiberiu Harko}
\email{tiberiu.harko@aira.astro.ro}
\affiliation{Department of Physics, Babeș-Bolyai University, Kogălniceanu Street, Cluj-Napoca, 400084, Romania}
\affiliation{Astronomical Observatory, 19 Cireșilor Street, Cluj-Napoca 400487, Romania.}
\author{Ujjal Debnath}
\email{ujjaldebnath@gmail.com} 
\affiliation{Department of
Mathematics, Indian Institute of Engineering Science and
Technology, Shibpur, Howrah-711 103, India}
\author{G. Mustafa}
\email{gmustafa3828@gmail.com}
\affiliation{Department of Physics,
Zhejiang Normal University, Jinhua 321004, People’s Republic of China}

\begin{abstract}
We consider the cosmological implications of a four-dimensional extension of the Gauss-Bonnet $f(\mathcal{G})$ gravity, where $\mathcal{G}$ is the Gauss-Bonnet topological invariant, in which the Einstein-Hilbert action is replaced by an arbitrary function $f(R,\mathcal{G},T)$ of $\mathcal{G}$, of the Ricci scalar $R$, and of the trace $T$ of the matter energy-momentum tensor. By construction, the extended Gauss-Bonnet type action involves a non-minimal coupling between matter and geometry. The field equations of the model are obtained by varying the action with respect to the metric. The generalized Friedmann equations, describing the cosmological evolution in the flat Friedmann-Lemaitre-Robertson-Walker geometry, are also presented in their general form.  We investigate the cosmological evolution of the Universe in the generalized Einstein-Gauss-Bonnet theiry for a specific choice of the Lagrangian density, as given by \(f(R,\mathcal{G},T) = \alpha_1 \mathcal{G}^{m} + \alpha_2 R^{\beta} - 2\alpha_3 \sqrt{-T},\) where \( \alpha_i \) (\( i = 1, 2, 3 \)), \( m \), and \( \beta \) are model parameters. First, the theoretical predictions of the model are compared with a set of observational data (Cosmic Chronometers, Type IA Supernovae, Baryon Acoustic Oscillations) via an MCMC analysis, which allows us to obtain constraints on the model parameters. A comparison with the predictions of the $\Lambda$CDM system is also performed. Next, the generalized Friedmann equations are reformulated as a dynamical system, and the properties of its critical points are studied by using the Lyapunov linear stability analysis. This investigation allows for the reconstruction of the Universe's history in this model, from the early inflationary era to the late accelerating phase. The statefinder diagnostic parameters for the model are also considered from the dynamical system perspective.  
\end{abstract}

\pacs{}
\maketitle
\tableofcontents

\section{Introduction}
The unexpected discovery of the late acceleration of the Universe \cite{acc1,acc2,acc3,acc4,acc5}, which showed that a transition from a decelerating evolution to an accelerating one did occur at a small redshift $z$ of the order of $z\approx 0.6$, led to the necessity of the reconsideration of the theoretical foundations of one of the most successful gravitational theories ever known, Einstein’s theory of general relativity. The simplest possibility of explaining the late, de Sitter type period of the Universes evolution is to resort again, and include in the field equations the cosmological constant $\Lambda$, introduced by Einstein in 1917 \cite{Ein}, with the main goal of building a static, general relativistic cosmological model. The cosmological constant had a complicated history \cite{Wein}, but presently the inclusion of $\Lambda$ in the gravitational field equations represents one of the theoretical basis of the present day standard cosmological paradigm, the $\Lambda$CDM model. The $\Lambda$CDM model also postulates the presence of another important (but still unobserved) component of the Universe, dark matter \cite{dark}. The $\Lambda$CDM model fits very well the cosmological data \cite{fit1,fit2,fit3,fit4}.

However, the $\Lambda$CDM paradigm is confronted with a major theoretical problem: no  physical theory that can explain it is known presently. The main problem has its origin when one tries to find the nature and the physical/geometrical interpretation of the cosmological constant \cite{Wein,Wein1a,Wein2a}. The physical interpretation of $\Lambda$ as representing the Planck-scale vacuum energy density $\rho_{vac}$ has led to the ``worst prediction
in physics" \cite{Lake}. The simplest possible estimation of the vacuum energy density gives
\bea
\rho _{vac}&\approx& \frac{\hbar}{c}\int _{k_{dS}}^{k_{Pl}}{\sqrt{k^2+\left(\frac{mc}{\hbar}\right)^2}d^3k}
\approx  \rho _{Pl}=\frac{c^5}{\hbar G^2}\nonumber\\
&=& 10^{93}\; {\rm \frac{g}{cm^3}}.
\eea
This result differs by a factor of around $10^{-120}$  from the observed value of the energy density associated to $\Lambda$, $\rho _{\Lambda}=\Lambda c^2/8\Xi G\approx 10^{-30}\;{\rm g/cm^3}$ \cite{C3}.
Moreover, at the observational level, the $\Lambda$CDM model is confronted with some very important challenges.  Perhaps the most important problem of the $\Lambda$CDM model is the so-called "Hubble tension", which has its origins in the important differences in the values of the present-day value of the Hubble function, $H_0$, determined from the CMB measurement \cite{fit4}, and obtained from the observations in the local Universe, at low redshifts \cite{M1,M2,M3,M4,M5,M6}. The determination of $H_0$ by the SHOES collaboration has obtained the value
$H_0 = 74.03 \pm 1.42$ km/s/Mpc \cite{M1}. However, the early Universe determinations, based on Planck satellite data, give the value $H_0 = 67.4 \pm 0.5$ km/s/Mpc \cite{C3}, a value that
deviates by $\sim  5\sigma$ from the result of SH0ES \cite{M5,M6}.

Moreover,  other fundamental theoretical problems cannot be answered within the $\Lambda$CDM model. Thus, the smallness of $\Lambda$, why it is so fine tuned, and why the transition from decelerating to the accelerating phases took place only recently are still open problems of cosmology \cite{Peeb}.  More importantly, the fundamental question if a cosmological constant is really needed to build up the foundations of observational and theoretical cosmology is still awaiting an answer.

Thus, the examination of alternative approaches for the understanding of the gravitational force could provide us with the possibility of finding a solution to the basic cosmological problems, without having to resort to a cosmological constant. In fact, there are (at least) three theoretical approaches that replace the cosmological constant $\Lambda$CDM in the basic gravitational theory, through the generalization of Einstein's standard gravitational theory.

The first important avenue for extending general relativity  is represented by the dark components model. It is based on the generalization of the Einstein field equations through the addition of one or two new terms to the standard matter energy-momentum tensor. The two new terms represent the dark energy,  the dark matter, respectively, or both. Therefore, in this approach, the gravitational interaction and dynamics are described by the generalized field equation \cite{HL20}
\be
G_{\mu \nu}=\kappa ^2 T^{ (mat)}_{\mu \nu}+\kappa ^2T^{ (DE)}_{\mu \nu}(\varphi, \psi _{\mu},...+\kappa ^2T^{ (DM)}_{\mu \nu}(\phi, \psi _{\mu},...)),
\ee
where by $G_{\mu \nu}$ we have denoted the Einstein tensor, $\kappa ^2$ is the gravitational coupling constant, and $T^ {(mat)}_{\mu \nu}$, $T^{(DE)}_{\mu \nu}(\phi, \psi_{\mu},...)$ and $T^{ (DM)}_{\mu \nu}(\varphi, \sigma_{\mu},...)$, and  denote the energy-momentum tensors of the baryonic matter, dark energy, and dark matter, respectively.  The energy-momentum tensors of dark energy and dark matter are obtained with the help of the scalar $\varphi$, $\phi$ or vector fields $\psi _{\mu}$, $\sigma _\mu$.  In the simplest dark components model dark energy is described by a scalar field $\phi$, with a self-interaction potential $V(\phi)$, with the gravitational action given by
\be\label{1}
S=\int{\left[\frac{\kappa ^2}{2}R-\partial _\mu\varphi \partial ^\mu \varphi-V(\varphi)\right]\sqrt{-g}d^4x}.
\ee

The dark energy models based on action (\ref{1}) are generally called quintessence models \cite{Qa1,Qa2,Qa3,Qa4,Qa5, Qa6,Qa7,Qa8}.  There are a large number of dark component models, like, for example,  k-essence models \cite{K1,K2,K3}, tachyon \cite{T1,T2}, phantom \cite{Ph1,Ph2,Ph3},  quintom \cite{Qu1,Qu2,Qu3} and chameleon \cite{Ch1,Ch2,Ch3,nonlocal,Ch4} field models.  The dark energy models of Chaplygin gas \cite{Cha1,Cha2} and vector field \cite{V1,V2,V3} have also been extensively used as dark energy models. Reviews of the dark energy models can be found in  \cite{Rev1,Rev2,Rev3,Rev4}.

The dark component models are remarkable successful in explaining cosmological dynamics. However,  they still face some intrinsic problems in relation with observations. Thus, for example, in \cite{Qa6} it was shown that in quintessence models the value of $H_0$ is always smaller compared to the $\Lambda$CDM model. Thus, in some quintessence models, the Hubble tension cannot be solved.

The second possible extension of general relativity is represented by the dark gravity type theories. Similarly to Einstein's formulation, dark gravity is based on a geometrical formalism for the description of gravity,  with the evolution of the Universe and the gravitational dynamics explained by the modification of the geometry of the space-time, which goes beyond the standard Riemannian geometry. The introduction of new geometrical structures also requires the generalization of the Hilbert-Einstein action.
In the dark gravity formalism, the Einstein field equations can be generally written down in the form
\be
G_{\mu \nu}=\kappa ^2T_{\mu \nu}^{(mat)}+\kappa ^2 T_{\mu \nu}^{((geom))}\left(g_{\mu \nu}, R, Q, \mathcal{T}, \square R,...\right),
\ee
where $T_{\mu \nu}^{(\rm geom)}\left(g_{\mu \nu}, R, Q,\mathcal{T}, \square R,...\right)$, denotes  an effective energy-momentum tensor, which can describe independently dark energy and dark matter, or both.  $T_{\mu \nu}^{((geom))}\left(g_{\mu \nu}, R, Q, \mathcal{T}, \square R,...\right)$, the effective energy-momentum tensor of dark energy (dark matter) is a purely geometric quantity, constructed generally with the help of the metric, of the Ricci scalar $R$, of the torsion $\mathcal{T}$, and of the nonmetricity $Q$, respectively.

Dark gravity type theories were first proposed within the framework of the $f(R)$ theory, initially introduced in \cite{Bu1}, and later further investigated in \cite{Bu2,Bu3,Bu4,Bu5}. In the $f(R)$ gravity theory, the Einstein-Hilbert action $S=\int{\left(R/\kappa ^2+L_m\right)\sqrt{-g}d^4x}$ of general relativity is modified into an action of the form $S= \int{\left[f(R)/\kappa^2+L_m\right]\sqrt{-g}d^4x}$, where $f(R)$ is an analytical function of the Ricci scalar $R$. The astrophysical and cosmological implications of the $f(R)$ type models have been extensively investigated \cite{fR1,fR2,fR3,fR4,fR6,fR7,fR8,fR9,fR10, fR11}.  

The first internally consistent $f(R)$ model was introduced in \cite{fR7}, with the corresponding model describing both the inflation and the late dark energy phases. Torsion can also be used to describe the gravitational interaction by replacing the curvature with torsion. The torsion based approach to gravity is known as the teleparallel equivalent of general relativity (TEGR) , which was extended to the $f(T)$ gravity theory \cite{fT1,fT2,fT3}. For a review of $f(T)$ gravity, see \cite{fT4}. Theories in which nonmetricity replaces curvature and allows the full description of the gravitational force have also been considered. They are known as $f(Q)$ type gravity theory \cite{fQ1,fQ2,fQ3,fQ4}. For a review of $f(Q)$ gravity, see \cite{fQ5}.

Another purely geometric theory, the hybrid Metric-Palatini Gravity Theory \cite{HMP1,HMP2,HMP3,HMP4}, generalizes and unifies two different geometric theories, the Palatini and the metric ones. The cosmological and astrophysical implications of the Weyl geometry have also been studied \cite{W1,W2,W3,W4,W5,W6,W7,W8,W9,W10}. For extensive reviews of dark gravity type theories, and of their applications, see \cite{R1,R2,R3,R4}.

A third possibility for the investigation of the gravitational force is represented by the dark coupling perspective.  In this approach, the basic idea consists in replacing the Einstein-Hilbert action, which has a simple additive structure in the curvature $R$ and $L_m$, where $L_m$ is the Lagrangian matter, with an arbitrary function of the geometric and matter terms. Hence, one can introduce a maximal extension of the Einstein-Hilbert action by assuming that generally the gravitational action can be represented by an analytical function of the curvature scalar $R$, of the torsion $\cal{T}$, non-metricity $Q$, matter Lagrangian $L_m$, trace $T$ of the energy-momentum tensor and other thermodynamic and geometric quantities. In the dark coupling avenue, a non-minimal coupling between matter and geometry naturally appears.

The Einstein gravitational field equations are generalized in the dark coupling approach to
\bea
G_{\mu \nu}&=&\kappa ^2T_{\mu \nu}^{(mat)}\nonumber\\
&&+\kappa ^2 T_{\mu \nu}^{(\rm coup)}\left(g_{\mu \nu},  R, Q,\mathcal{T},L_m, T, \square R, \square T,... \right).\nonumber\\
\eea

The effective energy-momentum tensor of the dark coupling type theories $T_{\mu \nu}^{(coup)}\left(g_{\mu \nu}, R, Q, \mathcal{T},L_m, T, \square R, \square T,... \right)$ is constructed from a  non-additive matter and geometric algebraic structure, which includes the couplings between all forms of baryonic matter, and the scalar and geometric quantities.
In \cite{fLm1} a gravitational theory with the action having the form
\be
S=\int{\left[f_1(R)+\left(1+\lambda f_2(R)\right)L_m\right]\sqrt{-g}d^4x},
\ee
was considered. This action was later generalized in \cite{fLm2}, and in  \cite{fLm3}, respectively, leading to the $f\left(R,L_m\right)$ gravity theory, with action $S=\int{f\left(R,L_m\right)\sqrt{-g}d^4x}$. The cosmological  and astrophysical implications of the $f\left(R,L_m\right)$ theory, as well as its theoretical aspects were considered in \cite{fLm4, fLm5,fLm6, fLm7,fLm8,fLm9,fLm10,fLm11}.

A different coupling between geometry and matter is considered in the $f(R,T)$ gravity theory \cite{fT1}, in which the Ricci scalar $R$ is non-minimally coupled to the trace of the matter energy-momentum tensor $T$.  The action of the theory is given by $S=\int{\left[f\left(R,T\right)+L_m\right]\sqrt{-g}d^4x}$. The astrophysical and cosmological implications of the theory of $f(R,T)$ were studied in \cite{fT2a,fT3a,fT4a,fT5a,fT6a,fT7a,fT8a,fT9a,fT10a,fT11a,fT12a}. A generalization of the $f(R,T)$ gravity theory was introduced in \cite{fT23},  with the action given by
\be
S=\frac{1}{16\Xi}\int{f\left(R,T,\Box T\right)\sqrt{-g}d^4x}+\epsilon \int{L_m\sqrt{-g}d^4x},
\ee
where $\epsilon =\pm 1$. Hence, the action is obtained by introducing higher derivative matter fields.

An extension of $f(\cal{T})$ gravity, allowing for a general coupling of the torsion scalar $\cal{T}$ with the trace of the matter energy-momentum tensor $T$ was considered in \cite{fTT}. The $f\left(\cal{T},T\right)$ theory represents a promising candidate for the explanation of the present day dynamics of the Universe. The coupling of geometry and matter in $f(Q)$ gravity was considered in \cite{fQC1,fQC2,fQC3}. For reviews discussing extensively the modified theories of gravity see \cite{RMG0, RMG1,RMG2,RMG3,RMG4}.

Among the modified theories of gravity of dark geometry type, the Gauss-Bonnet (GB) topological invariant modified gravity \cite{GB1, GB1a, GB2, GB3, GB4} has also attracted a lot of interest. The GB modified gravity theory is constructed by using the topological invariant called the GB invariant, usually denoted by $\mathcal{G}$, and which is an invariant in 4D space-time.
Since $\mathcal{G}$ is topologically invariant, in 4D it does not influence the dynamical behavior of particles and of the gravitational field in general relativity, when included in the Einstein-Hilbert action.

However, because of its interesting physical and mathematical properties, theoretical gravitational models including the GB invariant have been considered as potential candidates for explaining the recent observational data. There are three basic mathematical approaches that would allow the inclusion of the GB invariant into gravitational action. The first method implies the non-minimal coupling of $\mathcal{G}$ to a scalar field $\phi$  \cite{GB5,GB6, GB7}. The second avenue is represented by constructing the action as a nontrivial function of the topological invariant $f(\mathcal{G})$, an approach which is called the modified gravity theory $f(\mathcal{G}$). Thirdly, one can consider higher-dimensional theories in which $\mathcal{G}$ is no longer a topological invariant. After constructing the Einstein-Hilbert action with the GB term in a higher dimension $D$, and after taking the limit $D\rightarrow 4$, one obtains a non-trivial modified gravitational theory \cite{GB8}.

Cosmological and astrophysical constraints on the different versions of the GB modified gravity have been considered in \cite{GB9,GB10,GB11,GB12,GB13}. These studies in $f(\mathcal{G})$ gravity have pointed out its possibilities of mimicking the $\Lambda$CDM model, and to describe the radiation and matter dominated epochs, followed by an accelerating evolution \cite{GB14}. However, some problems related to the $f(\mathcal{G})$ have also been pointed out, and they are related to the presence of higher-order derivative terms in the gravitational field equations, which in the first-order perturbation models could lead to instabilities in the $f(\mathcal{G}$) theory \cite{GB15, GB16}. The criteria and conditions necessary to eliminate possible unphysical aspects of the theory have also been considered \cite{GB5}. For a review of the $f(\mathcal{G})$ theory see \cite{GB17}.

The coupling between the GB invariant and the matter term, described by the trace of the matter energy momentum tensor, was considered in \cite{GBn1}, which led to the formulation of the $f(R,\mathcal{G},T)$ gravity theory. Cosmological investigations of this theory were performed in \cite{GBn2,GBn3}. The gravitational baryogenesis in the framework of $f(R,\mathcal{G},T)$ gravity was considered in \cite{GBn3}, where constraints on the parameters of a specific model have been obtained.

It is the main goal of the present paper to further investigate the $f(R,\mathcal{G},T)$  extension of the $f(\mathcal{G})$ theory, which explicitly includes in the basic theory a non-minimal coupling between $\mathcal{G}$, the Ricci scalar $R$, and the trace of the matter energy-momentum tensor $T$.  The generalization of GB invariant modified gravity theory is realized by using the extension of the Einstein-Hilbert variational principle, as defined on a metric-affine manifold, endowed with the Levi-Civita connection. In the presence of matter on this manifold we consider three scalar quantities $(R,\mathcal{G},T)$,  from which two, $(R,\mathcal{G})$,  have a purely geometric origin, while $T$ describes the properties of the ordinary baryonic matter. The maximal extension of the Einstein-Hilbert action on the considered set of quantities is then represented by adopting for the gravitational Lagrangian density $L_g$ an arbitrary analytical function of the three scalars, so that $L_g=f(R,\mathcal{G}, T)$ \cite{GBn1}.  We assume, as is usual in modified gravity theories, that $f$, being an analytical function,  can be locally represented by a convergent power series. Furthermore, the Taylor series of $f$ about $x_0 =\left(R_0,\mathcal{G}_0,T_0\right)$ converges towards the function $f$ in some neighborhood for every $x_0$ in its domain. By construction, the action of the generalized GB theory $f(R,\mathcal{G},T)$ introduces a non-minimal coupling between geometry and matter, This coupling leads to the presence of a non-trivial gravitational dynamics, in which the matter component introduces a supplementary degree of freedom in the physical theory.

As a first step in our study, we write down the field equations of the $f(R,\mathcal{G},T)$ theory \cite{GBn1}, by varying the action with respect to the metric tensor. The cosmological implications of the theory are investigated by first presenting the generalized Friedmann equations for the case of the Friedmann-Lemaitre-Robertson-Walker (FLRW) metric. The equations of cosmological evolution explicitly depend on the derivatives of $f$ with respect to $\mathcal{G}$, $R$, and $T$. Then a particular cosmological model is investigated, obtained for a specific choice of the function $f$, assumed to have the form $f(R,\mathcal{G},T)=\alpha_1\mathcal{G}^m+\alpha_2R^{\beta}-2\alpha_3\sqrt{-T}$, where $m,\beta$ are constants.

 The implications of the introduced cosmological model are investigated from two distinct perspectives. First of all, the basic evolution equation for the Hubble function $H$ is obtained and studied numerically. The theoretical predictions of the model are then compared with several sets of observational data by using the MCMC (Markov Chain Monte Carlo) statistical procedure. This allows us to obtain strong observational constraints on the model parameters. A comparison with the predictions of the $\Lambda$CDM model is also performed. A second possibility of investigating a cosmological model is by the use of the methods of the theory of dynamical systems, and more exactly through the study of the Lyapunov stability of the critical points of the model \cite{dyns1,dyns2}. 
 
 After reformulating the generalized Friedmann equations as a first-order dynamical system by using a set of appropriate variables, the critical points of the system are obtained, and their Lyapunov stability is investigated. This analysis allows reconstructing the evolutionary history of the Universe in the framework of the $f(R,\mathcal{G},T$) Einstein- GB cosmological model with geometry matter coupling, which evolved from an inflationary era to a late-time accelerating phase, after passing through radiation- and matter-dominated epochs.
 
The present paper is organized as follows.  The action of the $f(R,\mathcal{G},T)$ theory is described in Section~\ref{sect1}, where the gravitational field equations are derived via the metric variation. The generalized Friedmann equations are also presented. A particular cosmological model, corresponding to a particular choice of the function $f$ is also introduced, and the Hubble function is obtained as a function of redshift and model parameters. The comparison of the cosmological model with several sets of observational data is performed in Section~\ref{sect2}, where strong constraints on the observational parameters are obtained.  The stability analysis of the system is performed in Section~\ref{sect3}, by reformulating the cosmological evolution equations as a dynamical system and obtaining its critical points. The statefinder parameters are considered from the dynamical system perspective in Section~\ref{sect4}. Finally, we discuss and conclude our results in Section~\ref{sect5}.
     
\section{Overview of $f\left(R,\mathcal{G},T\right)$ gravity}\label{sect1}

In four-dimensional 4D Riemannian geometry, there are two topological invariants, given by \cite{Anjos}
\begin{equation}
    I = \int \sqrt{- g} \, A \, d^{4}x,
\end{equation}
and
\begin{equation}
    II = \int \sqrt{- g} \, \mathcal{G} \, d^{4}x,
\end{equation}   
respectively,  where
\begin{equation}
    A = R_{\alpha\beta\mu\nu}^{*} \, R^{\alpha\beta\mu\nu},
\end{equation}    
and $\mathcal{G}$ is the GB topological invariant
\begin{equation}
    \mathcal{G} = R_{\alpha\beta}^{*}{}_{\mu\nu}^{*} \, R^{\alpha\beta\mu\nu}.
\end{equation}

The dual of any anti-symmetric tensor $F_{\mu\nu}$ is defined in the usual way as
\begin{equation} 
    F_{\mu\nu}^{*} = \frac{1}{2} \, \eta_{\mu\nu\alpha\beta} \, F^{\alpha\beta},
\end{equation}
where $\eta_{\mu\nu\alpha\beta} = \sqrt{-g} \, \epsilon_{\mu\nu\alpha\beta}$, and $ \epsilon_{\mu\nu\alpha\beta} $ is the completely anti-symmetric Levi-Civit\`a symbol. 
The invariant $ \mathcal{G} $ is obtained in terms of the curvature tensor and of its contraction by the identity
\begin{equation}
  \mathcal{G}=R^2-4R_{\mu\nu}R^{\mu\nu}+R_{\mu\nu\xi\eta}R^{\mu\nu\xi\eta}.
\end{equation}

The Einstein-Hilbert action for $f(R,\mathcal{G},T)$ gravity theory is defined as  \cite{GBn1}
\begin{eqnarray}\label{eq2}
    S=\frac{1}{2}\int{f(R,\mathcal{G},T)\sqrt{-g} d^4x}+\int{L_M}\sqrt{-g}d^4x.
\end{eqnarray}

Here $f(R,\mathcal{G},T)$ is an arbitrary analytic function of the Ricci scalar $R=g^{\mu\nu}R_{\mu\nu}$, of the GB invariant $\mathcal{G}$, and of the trace of the matter energy-momentum tensor $T=g^{\mu \nu}T_{\mu \nu}$. $L_M$ is the Lagrangian density corresponding to the matter sector. Now, using the least action principle, from Eq.~\eqref{eq2}, we obtain the field equation as \cite{GBn1}
\begin{widetext}
\begin{eqnarray}\label{eq3}
    &&\left(R_{\mu\nu}+g_{\mu\nu}\Box-\nabla_\mu \nabla_\nu\right)f_R-\frac{1}{2}f g_{\mu\nu}+\left(2RR_{\mu\nu}-4R_\mu^\xi R_{\xi \nu}-4R_{\mu\xi\nu\eta}R^{\xi\eta}+2R_{\mu}^{\xi\eta\lambda}R_{\nu\xi\eta\lambda}\right)f_{\mathcal{G}}\nonumber\\
    &&+\left(2Rg_{\mu\nu}\Box-2R\nabla_\mu \nabla_\nu-4g_{\mu\nu}R^{\xi\eta}\nabla_\xi \nabla_\eta-4R_{\mu\nu}\Box+4R^\xi_{\mu} \nabla_\nu \nabla_\xi+4R^\xi_{\nu}\nabla_{\mu}\nabla_\xi+4R_{\mu\xi\nu\eta}\nabla^\xi \nabla^\eta\right)f_\mathcal{G} \nonumber\\
   &&=T_{\mu\nu}-\left(T_{\mu\nu}+\Theta_{\mu\nu}\right)f_T,
\end{eqnarray}
\end{widetext}
where we have denoted $f_R=\frac{\partial f}{\partial R}$, $f_\mathcal{G}=\frac{\partial f}{\partial \mathcal{G}}$, $f_T=\frac{\partial f}{\partial T}$, $\Box=\nabla^2=\nabla_\mu\nabla^\mu$, and 
\be\label{theta}
\Theta_{\mu\nu}=-2T_{\mu\nu}+pg_{\mu\nu}.
\ee
Taking the trace of the above field equation and multiplying by $g^{\mu\nu}$, we get
\bea\label{eqq17}
  \left(R^2+3\Box\right)f_R&-&2f-\left(2\mathcal{G}-2R \Box+4R^{\mu\nu}\nabla_\mu \nabla_\nu\right)f_\mathcal{G}\nonumber\\
  &=&T-\left(T+\Theta\right)f_T ,
\eea
where we have denoted $\Theta=\Theta_{\mu\nu}g^{\mu\nu}$.

By taking the covariant divergence of the equation ~(\ref{eqq17}), we find the relation
\begin{eqnarray}\label{eqq18}
    \nabla^\mu T_{\mu\nu}&=&\frac{f_T}{1-f_T} \Bigg[\left(T_{\mu \nu}+\Theta_{\mu\nu}\right)\nabla^\mu \ln{f_T}+\nonumber\\
    &&\nabla^\mu \Theta_{\mu\nu}-\frac{1}{2}g_{\mu\nu}\nabla^\mu T \Bigg].
\end{eqnarray}
In Eq.~(\ref{theta})  $T_{\mu\nu}$ is the energy-momentum tensor for a perfect fluid, given by 
\begin{eqnarray}
    T_{\mu\nu}=\left(\rho+p\right)u_\mu u_\nu+pg_{\mu\nu},
\end{eqnarray}
where $\rho$ and $p$ are the energy density and the pressure, respectively, and $u_\mu$ is the four-velocity vector that satisfies the conditions $u_\mu u^\mu=-1$ and $u^\mu \nabla_\nu u_\mu=0$, respectively.

\subsection{The generalized Friedmann equations}
We now consider the cosmological applications of the $f(R,\mathcal{G},T)$ theory. We assume that the background is the flat, homogenous, and isotropic FLRW geometry, with metric given by 
\begin{eqnarray}\label{eq1}
 ds^2=-dt^2 +a^2(t) \left(\frac{dr^2}{1-K r^2}+r^2 d\Omega^2 \right) ,
\end{eqnarray}
where $ d\Omega^2=d\theta^2 +\sin{\theta}^2 d\phi^2$. The expressions for the Ricci scalar $R$ and for the GB invariant $G$ can be written as 
\begin{eqnarray}\label{eq5}
    R=6\left(\dot{H}+2H^2\right), \mathcal{G}=24H^2\left(\dot{H}+H^2\right).
\end{eqnarray}
Here, $H=\frac{\dot{a}}{a}$ is the Hubble parameter, with a dot representing the derivative with respect to cosmic time $t$. Also for the metric \eqref{eq1}, the trace of the energy-momentum tensor $T_{\mu\nu}$ is $T=3p-\rho$.

Now from the equation ~(\ref{eqq18}), we obtain the non conservation equation of the matter energy momentum tensor as
\begin{equation}\label{non conservation}
   \dot{\rho}+3H\left(\rho+p\right)=\left(\frac{1}{2}\dot{T}-\dot{p}\right)f_T -\left(\rho+p\right)\dot{f_T}. 
\end{equation}
In the following, we will investigate a conservative model, in which the energy-momentum tensor is divergence less, i.e., $\nabla_\mu T^{\mu\nu}=0$. This assumption leads us to obtain the baryonic matter conservation equation as
\begin{eqnarray}\label{conservation eqn}
    \dot{\rho} +3H\left(\rho+p\right)=0.
\end{eqnarray}
From Eqs.~(\ref{non conservation}) and ~(\ref{conservation eqn}), we have 
\begin{equation}\label{24}
    \left(\dot{\rho}-\dot{p}\right)f_T+2\left(\rho+p\right)\dot{f_T}=0.
\end{equation}
By using Eq.~\eqref{eq3}, we obtain the cosmological field equations (the generalized Friedmann equations) for the $f\left(R,\mathcal{G},T\right)$ gravity as 
\begin{eqnarray}\label{eq7}
    3H^2&=&\frac{1}{f_R}\Bigg[\rho +\left(\rho+p\right)f_T+\frac{1}{2}\left(Rf_R -f\right)-3H\dot{f_R}\nonumber\\
    &&+12H^2\left(\dot{H}+H^2\right)f_\mathcal{G}-12H^3\dot{f_\mathcal{G}}\Bigg],
\end{eqnarray}
\begin{eqnarray}\label{2nd friedmann}
    2\dot{H}+3H^2&=&-\frac{1}{f_R}\Bigg[p-\frac{1}{2}\left(Rf_r-g\right)+2H\dot{f_R}+\Ddot{f}R \nonumber\\
    &&-12H^2+\left(\dot{H}+H^2\right)f_\mathcal{G}+8H\left(\dot{H}+H^2\right)\dot{f}_\mathcal{G}\nonumber\\
    &&+4H^2\Ddot{f}_\mathcal{G}\Bigg].
\end{eqnarray}
One can always write down the above generalized Friedmann equations in the conventional form of the standard Friedmann  evolution equations
\begin{eqnarray}
    3H^2&=&\rho_{eff},\\
    2\dot{H}+3H^2&=&-p_{eff},
\end{eqnarray}
with the  respective expression corresponding to $\rho_{eff}$ and $p_{eff}$ being given by
\begin{eqnarray}
    \rho_{eff}&=&\frac{1}{f_R}\Bigg[\rho +\left(\rho+p\right)f_T+\frac{1}{2}\left(Rf_R -f\right)-3H\dot{f_R}\nonumber\\
    &&+12H^2\left(\dot{H}+H^2\right)f_\mathcal{G}-12H^3\dot{f_\mathcal{G}}\Bigg],
\end{eqnarray}
and 
\begin{eqnarray}
  p_{eff}&=&\frac{1}{f_R}\Bigg[p-\frac{1}{2}\left(Rf_R-f\right)+2H\dot{f_R}+\Ddot{f}R-12H^2\nonumber\\
  &&+\left(\dot{H}+H^2\right)f_G+8H\left(\dot{H}+H^2\right)\dot{f}_\mathcal{G}+4H^2\Ddot{f}_\mathcal{G}\Bigg],\nonumber\\
\end{eqnarray}
respectively.

In the present work, we assume that the Universe is filled with pressureless dust, and therefore we set $p=0$. One may note that all the background equations are highly sensitive to the choice of functional $f$. 

In order to proceed in our study on the observational data and analysis of the dynamical system of the $f\left(R,\mathcal{G},T\right)$ gravity theory, it is important to consider a specific form of the functional $f$ that satisfies all the fundamental conditions of the theory. 

Since we have assumed baryonic matter conservation, and the pressureless dust condition, Eq.~ (\ref{24}) gives for $f_T$ the equation
\be
\dot{\rho}f_T+2\rho \dot{f}_T=0,
\ee
with the general solution given by
\be
f_T=\frac{C}{\sqrt{\rho}}=\frac{C}{\sqrt{-T}},
\ee
where $C$ is an arbitrary integration constant. Hence, by further integration with respect to $T$ we find
\be
f(R,\mathcal{G},T)=-2\alpha_3\sqrt{-T}+\xi(R,\mathcal{G}),
\ee
where $\xi(R,\mathcal{G})$ is an arbitrary integration function, and we have redefined the integration constant $C$.  As for the function $\xi(R,\mathcal{G})$ we assume the form $\xi\left(R,\mathcal{G}\right)=\alpha_1 \mathcal{G}^m + \alpha_2 R^{\beta} $, and thus the present study we consider for the function $f$  the form 
\begin{eqnarray}
f(R,\mathcal{G},T) = \alpha_1 \mathcal{G}^{m} + \alpha_2 R^{\beta} - 2\alpha_3 \sqrt{-T},
\end{eqnarray}
Where \( \alpha_i \), \( i = 1, 2, 3 \), are coupling parameters, describing the interaction strength of the Gauss-Bonnet invariant \( \mathcal{G} \), the Ricci scalar \( R \), and the trace of the energy-momentum tensor \( T \) , respectively and $m$, \( \beta \) are the model parameters that characterizes the non-linear effect of the topological Gauss-Bonnet invariant $\mathcal{G}$ and the Ricci scalar \( R \) respectively, in the gravitational sector.

The expression for the Hubble parameter $H$ in terms of the redshift $z$, and the other cosmological parameters is obtained as a solution of the following second order ordinary differential equation
\begin{widetext}
{\scriptsize
\begin{eqnarray}\label{hubble_full}
H''(z) &=& -3 H(z)^2 - \frac{9 (-1 - z) (-1 + \beta) H(z)^2 H'(z)}{2 H(z) - (1 + z) H'(z)} + \frac{3 (-1 - z) (1 + z) (-1 + \beta) H(z) H'(z)^2}{2 H(z) - (1 + z) H'(z)} \nonumber \\
&& + 3 H(z) \left(2 H(z) - (1 + z) H'(z)\right) - \frac{3 H(z) \left(2 H(z) - (1 + z) H'(z)\right)}{\beta} \nonumber \\
&& + \frac{2^{1 - \beta} 3^{2 - \beta} H_0^2 (1 + z)^3 \Omega_{\text{m0}} \left(H(z)\left(2 H(z) - (1 + z) H'(z)\right)\right)^{1 - \beta}}{\beta \alpha_2} \nonumber \\
&& + \frac{2^{2 - \beta} 3^{\frac{3}{2} - \beta} \alpha_3 \sqrt{H_0^2 (1 + z)^3 \Omega_{\text{m0}}} \left(H(z)\left(2 H(z) - (1 + z) H'(z)\right)\right)^{1 - \beta}}{\beta \alpha_2} \nonumber \\
&& - \frac{2^{3 m - \beta} 3^{1 + m - \beta} \alpha_1 \left(H(z)^3 \left(H(z) - (1 + z) H'(z)\right)\right)^m \left(H(z)\left(2 H(z) - (1 + z) H'(z)\right)\right)^{1 - \beta}}{\beta \alpha_2} \nonumber \\
&& + \frac{2^{3 m - \beta} 3^{1 + m - \beta} m \alpha_1 \left(H(z)^3 \left(H(z) - (1 + z) H'(z)\right)\right)^m \left(H(z)\left(2 H(z) - (1 + z) H'(z)\right)\right)^{1 - \beta}}{\beta \alpha_2} \nonumber \\
&& + \frac{2^{3 m - \beta} 3^{2 + m - \beta} (-1 + m) m (1 + z) H(z) \alpha_1 H'(z) \left(H(z)^3 \left(H(z) - (1 + z) H'(z)\right)\right)^m \left(H(z)\left(2 H(z) - (1 + z) H'(z)\right)\right)^{1 - \beta}}{\beta \alpha_2 \left(H(z) - (1 + z) H'(z)\right)^2} \nonumber \\
&& - \frac{2^{3 m - \beta} 3^{2 + m - \beta} (-1 + m) m (1 + z)^2 \alpha_1 H'(z)^2 \left(H(z)^3 \left(H(z) - (1 + z) H'(z)\right)\right)^m \left(H(z)\left(2 H(z) - (1 + z) H'(z)\right)\right)^{1 - \beta}}{\beta \alpha_2 \left(H(z) - (1 + z) H'(z)\right)^2} \nonumber \\
&& \Bigg/ \left[ - \frac{3 (-1 - z) (1 + z) (-1 + \beta) H(z)^2}{2 H(z) - (1 + z) H'(z)} \right. \nonumber \\
&& \left. + \frac{2^{3 m - \beta} 3^{1 + m - \beta} (-1 + m) m (1 + z)^2 H(z) \alpha_1 \left(H(z)^3 \left(H(z) - (1 + z) H'(z)\right)\right)^m \left(H(z)\left(2 H(z) - (1 + z) H'(z)\right)\right)^{1 - \beta}}{\beta \alpha_2 \left(H(z) - (1 + z) H'(z)\right)^2} \right] \nonumber \\
\end{eqnarray}
}
\end{widetext}
where \( H = H(z) \), \( H' = \frac{dH(z)}{dz} \), and \( H'' = \frac{d^2H(z)}{dz^2} \) represent the Hubble parameter and its first and second derivatives with respect to the redshift \( z \), respectively.

In the following, we denote by \( H_0 \) the current value of the Hubble parameter (i.e., the Hubble constant), \( \Omega_{m0} \) denotes the present-day matter density parameter and \( \alpha_i \), \( i = 1, 2, 3 \), $m$, and \( \beta \) are model parameters.

\section{Comparison with the observational data}\label{sect2}

In this Section, we obtain the posterior distribution of the parameters of the \( f(R, \mathcal{G}, T) \) model as described by the Hubble function \eqref{hubble_full}. To achieve this, we solve the second-order differential equation governing the Hubble parameter \( H(z) \), given by Eq.~(\ref{hubble_full}). The equation incorporates key cosmological parameters, including the Hubble constant \( H_0 \), the present-day matter density parameter \( \Omega_{m0} \), and \( \alpha_i \) (for \( i = 1, 2, 3 \)), $m$, and \( \beta \) are model parameters. 

Given the complexity of the equation, we employ numerical methods for its solution. We reformulate the second-order differential equation into a system of first-order equations by introducing \( u = dH/dz \) as an auxiliary variable. This allows us to express the system as
\be
\frac{dH}{dz} = u,
\ee
and
\be
\frac{du}{dz} = f(H, u, z, \alpha_1, \alpha_2, \alpha_3, \beta, m),
\ee
respectively, where \( f(H, u, z, \alpha_1, \alpha_2, \alpha_3, \beta, m) \) denotes the right-hand side of the governing equation, Eq.~(\ref{hubble_full}). 

The initial conditions are set at redshift \( z = 0 \) as \( H(0) = H_0 \) and \( H'(0) = \frac{3}{2} H_0 \Omega_{m0} \), based on standard cosmological assumptions. To solve this system, we employ the  \textit{Runge Kutta (RK45)} method using the \textit{solve\_ivp} function from the \textit{SciPy library}. This adaptive step solver is well suited for stiff and non-stiff ordinary differential equations, ensuring accuracy while optimizing computational efficiency. The integration spans from \( z = 0 \) to \( z = 3 \), with a resolution of 800 points to ensure sufficient precision. The relative and absolute tolerances are set to \( 10^{-3} \) and \( 10^{-6} \), respectively, to balance numerical stability and computational cost. The numerical solution obtained provides the evolution of \( H(z) \), which is subsequently used to compute the likelihood function for parameter estimation. 

\subsection{Methodology and data description}
After obtaining the numerical solution of the second-order nonlinear Hubble function describing the cosmological evolution of the $f(R,\mathcal{G},T)$ gravity model, we estimate the posterior distribution of the cosmological parameters using the nested sampling algorithm, implemented via the \texttt{PolyChord} package\footnote{\protect\url{https://github.com/PolyChord/PolyChordLite}}. Bayes' theorem expresses the posterior as: $P(\theta | D) = \frac{P(D|\theta) P(\theta)}{P(D)}$ where $P(\theta | D)$ is the posterior, $P(D|\theta)$ is the likelihood, $P(\theta)$ is the prior, and $P(D)$ is the evidence. PolyChord uses nested sampling to efficiently explore the parameter space and calculate the evidence. This approach shrinks the prior volume while focusing on regions with higher likelihood, ensuring efficient sampling and an accurate posterior distribution. In contrast to traditional MCMC,  nested sampling directly computes the posterior and the Bayesian evidence in a more efficient manner. We use the PolyChord sampler, which provides posterior samples and evidence automatically, handling high-dimensional parameter spaces and model comparison efficiently. To visualize posterior distributions and analyze parameter correlations, we use \texttt{GetDist}\footnote{\protect\url{https://getdist.readthedocs.io/en/latest/plot_gallery.html}}, a Python library designed to handle and plot MCMC samples. \texttt{GetDist} generates detailed plots, including marginalized distributions and parameter correlation plots, which allow for effective visualization of the uncertainty and dependencies of the cosmological parameters within the $f\left(R,\mathcal{G},T\right)$ gravity model \cite{GetDist}.
\begin{itemize}
    \item \textbf{Cosmic Chronometers}: We consider a compilation of 15 Hubble measurements spanning from the redshift range $0.1791 \leq z \leq 1.965$, compiled from previous studies \cite{moresco2012new,moresco2015raising,moresco20166}. These measurements are derived using the differential age technique, which involves comparing passively evolving galaxies at nearby redshifts. By estimating the change in redshift with respect to cosmic time ($\Delta z / \Delta t$), this method provides a model-independent approach to determining the expansion rate of the Universe \cite{jimenez2002constraining}. To incorporate this dataset, we adopt the likelihood function provided by Moresco in his GitLab repository \footnote{\protect\url{https://gitlab.com/mmoresco/CCcovariance}}, which includes the full covariance matrix, incorporating both statistical and systematic uncertainties \cite{moresco2018setting,moresco2020setting}.
    \item \textbf{Type Ia Supernovae}: We consider the SNe Ia dataset without the SHOES calibration \cite{Brout2022}. The chi-square statistic for this dataset is given by the following expression: $\chi^2_{\text{SNe Ia}} = \Delta \mathbf{D}^T \mathbf{C}^{-1}_{\text{total}} \Delta \mathbf{D},$ where \( \Delta \mathbf{D} \) represents the differences between the observed distance moduli, \( \mu(z_i) \), and the model predictions, \( \mu_{\text{$f\left(R,\mathcal{G},T\right)$ Model}}(z_i, \theta) \) at redshift $z_i$ ( calculated for $f\left(R,\mathcal{G},T\right)$ model parameters $\theta$) \cite{Astier2006}. The differences are calculated as: $\Delta D_i = \mu(z_i) - \mu_{\text{$f\left(R,\mathcal{G},T\right)$ Model}}(z_i, \theta).$ The total covariance matrix, \( \mathbf{C}_{\text{total}} \), combines both statistical uncertainties (\( \mathbf{C}_{\text{stat}} \)) and systematic uncertainties (\( \mathbf{C}_{\text{sys}} \)) \cite{Conley2010}. Its inverse, \( \mathbf{C}^{-1}_{\text{total}} \), is used in the calculation. The model predicted distance modulus, \( \mu_{\text{$f\left(R,\mathcal{G},T\right)$ model}}(z_i) \), is given by: $\mu_{\text{model}}(z_i) = 5 \log_{10} \left( \frac{d_L(z)}{\text{Mpc}} \right) + \mathcal{M} + 25,$ where \( d_L(z) \) is the luminosity distance in a flat FLRW Universe, expressed as: $d_L(z) = c(1 + z) \int_0^z \frac{dz'}{H(z')}.$ Here, \( c \) is the speed of light in vaccum, and \( H(z) \) is the Hubble parameter. This method shows the degeneracy between the parameters \(\mathcal{M}\) and \(H_0\) when analyzing SNe Ia data.
    \item \textbf{Baryon Acoustic Oscillations}: We use the 13 Baryon Acoustic Oscillations (BAO) measurements from the Dark Energy Spectroscopic Instrument (DESI) Data Release 2 (DR2) \footnote{\url{https://github.com/CobayaSampler/bao_data}} \cite{Karim2025} spanning from the redshift range $0.295 \leq z \leq 2.330$. These measurements rely on the sound horizon at baryon decoupling, which is approximately \( r_d \approx 147.09 \pm 0.26 \, \text{Mpc} \) \cite{fit4}. To ensure model independence approach, \( r_d \) is treated as a free parameter, allowing constraints on \( r_d \) without assumptions about the physics of recombination or the early Universe \cite{Pogosian2020,Jedamzik2021,Pogosian2024,Lin2021,Vagnozzi2023}. To analyze the BAO dataset, we calculate the following key cosmological distances under the assumption of a flat Universe: the Hubble distance, Hubble distance $D_H(z)$, comoving angular diameter distance $D_M(z)$, and volume-averaged distance $D_V(z)$. To compare with cosmological models, we use the ratios $D_H(z)/r_d$, $D_M(z)/r_d$, and $D_V(z)/r_d$. The chi-squared statistic for the BAO measurements is defined as $\chi^2_i = \boldsymbol{\Delta}_i^{T} \mathbf{C}_i^{-1} \boldsymbol{\Delta}_i,$ where $\boldsymbol{\Delta}_i = \boldsymbol{\mathcal{O}}_i^{\text{th}} - \boldsymbol{\mathcal{O}}_i^{\text{obs}}$ represents the vector of residuals between theoretical and observed quantities, and $\mathbf{C}_i$ is the full covariance matrix\footnote{\url{https://github.com/CobayaSampler/bao_data/blob/master/desi_bao_dr2/desi_gaussian_bao_ALL_GCcomb_cov.txt}}.
\end{itemize}
The total chi-squared statistic, \(\chi^2_{\text{Tot}}\) for the model is given by the sum of the contributions from all datasets
\[
\chi^2_{\text{Tot}} = \chi^2_{\text{CC}} + \chi^2_{\text{SNe Ia}} + \chi^2_{\text{BAO}},
\]
Next, we computed the Bayes Factor, defined as: $B_{ij} = \frac{p(d|M_i)}{p(d|M_j)}$ where $p(d|M_i)$ represents the Bayesian evidence for model $M_i$, and $p(d|M_j)$ represents the evidence for model $M_j$. The Bayesian evidence is challenging to compute analytically, but in the PolyChord, it is computed numerically by algorithm. We use the natural logarithm of $B_{0i}$, where the model $M_0$ corresponds to the standard $\Lambda$CDM model, and we compare it against other models, denoted by the index $i$. Based on Jeffreys' scale \cite{Jeffreys1998}, the interpretation of $\ln(B_{ij})$ is as follows: $\ln(B_{ij}) < 1$ indicates that the evidence is inconclusive for either of the models. $1 \leq \ln(B_{ij}) < 2.5$ suggests weak support for model $i$ relative to model $j$. $2.5 \leq \ln(B_{ij}) < 5$ corresponds to moderate support for model $i$ over model $j$. $\ln(B_{ij}) > 5$ provides strong evidence in favor of model $i$ compared to model $j$. Conversely, if $\ln(B_{ij})$ is negative, it implies that the evidence favors model $j$ instead of model $i$, following the same scale.

\subsection{Observational and Statistical Results}
\begin{figure*}
\centering
\includegraphics[width=18.0 cm]{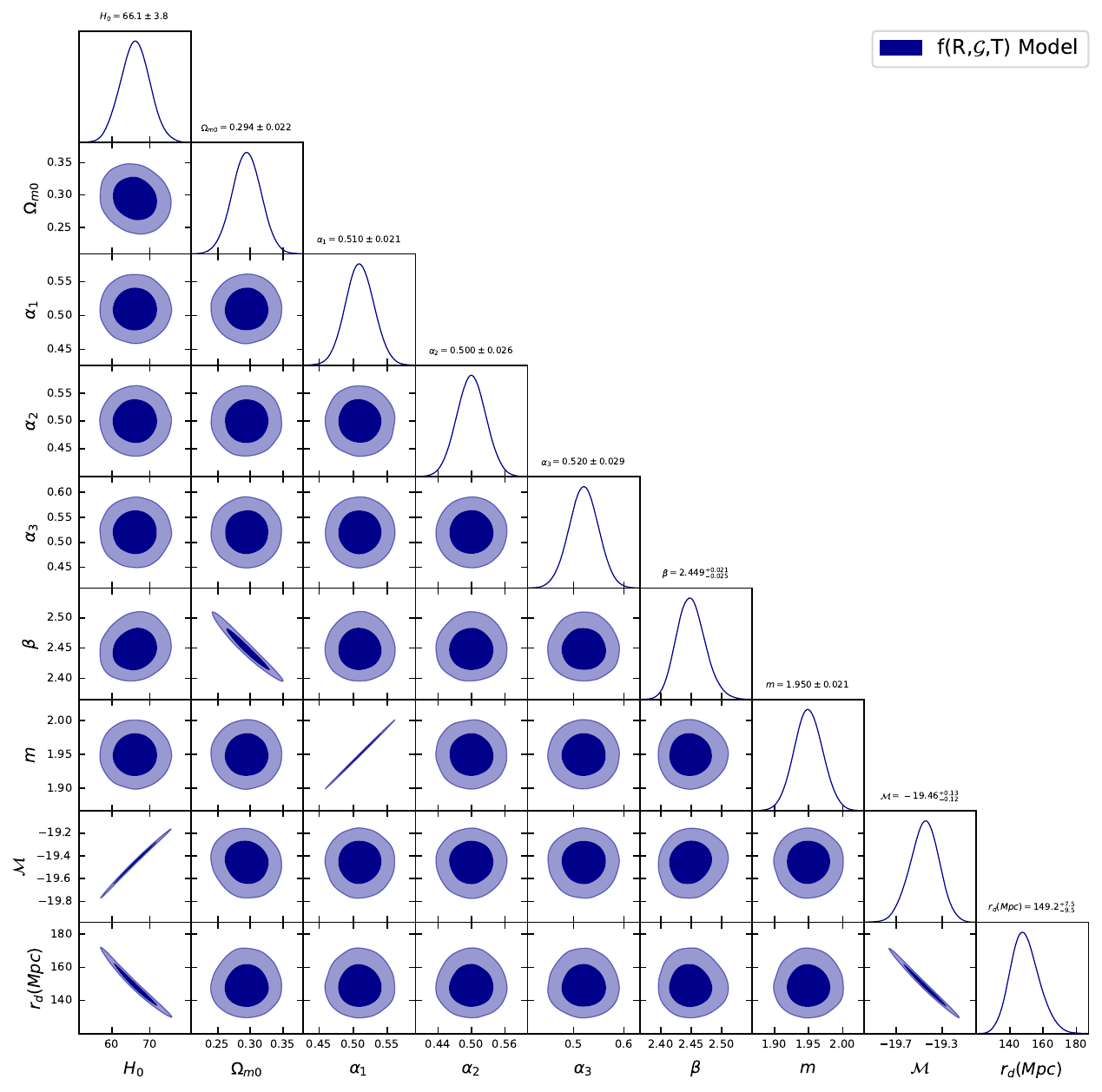}
\caption{The posterior distributions of the parameters of the \( f(R,\mathcal{G},T) \) model at 68\% (1\(\sigma\)) and 95\% (2\(\sigma\)) credible intervals.}\label{fig_1}
\end{figure*}

Fig.~\ref{fig_1} shows the triangle plot, which represents the posterior distributions of the \( f(R,\mathcal{G},T) \) model parameters. The diagonal panels show the marginalized probability distributions, where the peak represents the most probable value, and the spread indicates uncertainty. The off-diagonal panels display 2D contour plots, depicting parameter correlations, with the inner and outer contours denoting 68\% (1\(\sigma\)) and 95\% (2$\sigma$) confidence intervals, respectively. The figure highlights a significant negative correlation between $\Omega_{m0}$ and $\beta$.

Table~\ref{tab_1} presents the mean values, along with their 68\% (1$\sigma$) confidence intervals for the $\Lambda$CDM and $f(R,\mathcal{G},T)$ models. In both cases, the predicted values of $H_0$ are close to Moresco's results, considering both the full covariance matrix for the $\Lambda$CDM model and that the values of $r_d$ are consistent with those reported by the Planck Collaboration. For the $\Lambda$CDM model, the predicted values of $\Omega_{m0}$ and $\Omega_{\Lambda0}$ are closely aligned with the Planck results ($\Omega_{m0} = 0.315 \pm 0.007$, $\Omega_{\Lambda0} = 0.685 \pm 0.007$). However, in the case of the $f(R,\mathcal{G},T)$ model, the predicted value of $\Omega_{m0}$ is lower than the Planck estimate but is consistent with the value predicted by the DESI DR2 results ($\Omega_{m0} = 0.294 \pm 0.022$).

We observe that the central values of $H_0$ and $r_d$ are close to those predicted by the Planck Collaboration. Although the associated uncertainty is larger, this is primarily due to the use of the full cosmic covariance matrix, which incorporates uncertainties from stellar metallicity, star formation history, the Initial Mass Function (IMF), stellar libraries, synthesis models, and possible residual young stellar components.

Based on the statistical results, the $f(R, \mathcal{G}, T)$ model shows moderate support over the $\Lambda$CDM model, with a Bayes Factor value of $\ln(B_{0i}) = 3.5994$. This suggests that, while there is moderate evidence in favor of the $f(R, \mathcal{G}, T)$ model, the difference is not large enough to strongly rule out $\Lambda$CDM. Both models are viable, but the $f(R, \mathcal{G}, T)$ model is marginally favored by the data.
\begin{table}
\begin{tabular}{|c|c|c|c|}
\hline
Models & Parameter & Prior & JOINT  \\
\hline
& $H_{0}$ & $[50.,100.]$ & $67.9{\pm 4.0}$  \\[0.1cm]
& $\Omega_{m0}$ &$[0.,1.]$   & $0.3094 {\pm 0.0087}$  \\[0.1cm]
$\Lambda$CDM Model & $\mathcal{M}$ & $[-20,-18]$  &$-19.43{\pm 0.12}$  \\[0.1cm]
&$r_{d}$(Mpc)   &$[100.,300.]$  &$148.3{\pm 7.5}$  \\[0.1cm]
& $\Omega_{\Lambda0}$ &$[0.,1.]$   & $0.6906{\pm 0.0087}$ 
\\[0.1cm]
\hline
& $H_{0}$ & $[50.,100.]$ & $66.1{\pm 3.8}$  \\[0.1cm]
& $\Omega_{mo}$ &$[0.,1.]$   &$0.294{\pm 0.022}$ \\[0.1cm]
& $\alpha_1$ &$[0.,1.]$   &$0.510{\pm 0.021}$ \\[0.1cm]
$f\left(R,\mathcal{G},T\right)$ Model & $\alpha_2$ &$[0.,1.]$   &$0.500{\pm 0.026}$ \\[0.1cm]
& $\alpha_3$ &$[0.,1.]$   &$0.520{\pm 0.029}$ \\[0.1cm]
& $\beta$ &$[2.,3.]$   &$2.544{\pm 0.021}$ \\[0.1cm]
& $m$ &$[1.,3.]$   &$1.950{\pm 0.021}$ \\[0.1cm]
& $\mathcal{M}$ & $[-20.,-18.]$  &$-19.46{\pm 0.12}$  \\[0.1cm]
&$r_{d}$(Mpc)   &$[100.,300.]$  &$149.2{\pm 7.5}$  \\[0.1cm]
\hline
\end{tabular}
\caption{Mean values, along with 68\% (1\(\sigma\)) credible intervals, and prior ranges for the standard $\Lambda$CDM model and the \( f(R,\mathcal{G},T) \) model}\label{tab_1}
\end{table}
\subsection{Hubble Parameter $H(z)$} 
In this subsection, we plot the evolution of the Hubble parameter $H(z)$ and compare the predictions of the $f(R, \mathcal{G}, T)$ gravity model with the standard $\Lambda$CDM model. For the $\Lambda$CDM model, we use the following expression for the Hubble parameter: $H_{\Lambda \text{CDM}}(z) = H_0 \sqrt{\Omega_{m0}(1 + z)^3 + \Omega_{\Lambda0}},$ where $\Omega_{m0} = 0.309$, $\Omega_{\Lambda0} = 0.690$, and $H_0 = 67.9\, \mathrm{km\,s^{-1}\,Mpc^{-1}}$. For the $f(R, \mathcal{G}, T)$ gravity model, we solve the second-order nonlinear differential equations numerically to obtain the Hubble function. The resulting Hubble parameter is then compared to the predictions of the $\Lambda$CDM model. In addition to the Hubble parameter comparison, we also include the 1 $\sigma$ (68\%) and 2 $\sigma$ (95\%) confidence regions to quantify the uncertainty in the predictions of both models.

\begin{figure}
\centering
\includegraphics[width=8.5 cm]{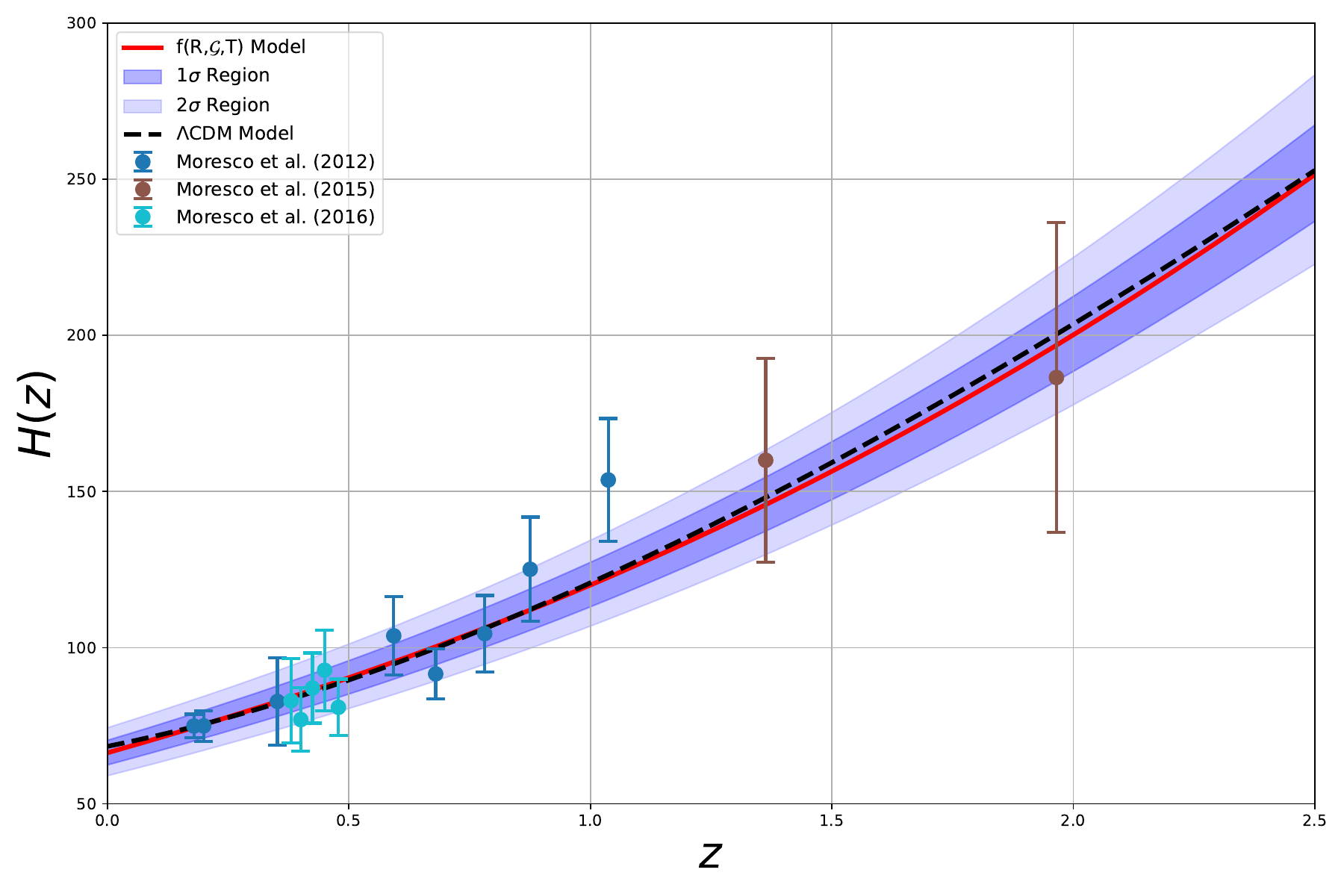}
\caption{Evolution of the Hubble parameter $H(z)$. A comparison of the $f(R, \mathcal{G}, T)$ gravity model with $\Lambda$CDM, using CC measurements. The analysis shows results within the 68\% (1$\sigma$) and 95\% (2$\sigma$) confidence regions.}\label{fig_2}
\end{figure}

Fig.~\ref{fig_2} shows the evolution of the Hubble function for the $f(R, \mathcal{G}, T)$ gravity model and the $\Lambda$CDM model. We observe that the $f(R, \mathcal{G}, T)$ gravity model aligns closely with the $\Lambda$CDM model at lower redshifts. However, as the redshift increases, small deviations begin to emerge between the two models. These differences, though present, are not significant. Additionally, the predictions of the $\Lambda$CDM model fall within the 1-$\sigma$ region of the $f(R, \mathcal{G}, T)$ gravity model, indicating that statistically, there is little to no significant difference between the two models.
\section{Stability analysis}\label{sect3}
In the present Section we will consider the analysis of the cosmological model described by the evolution Eqs.~\eqref{eq7} and \eqref{2nd friedmann} by using methods from the theory of the dynamical systems. In particular, the linear (Lyapounov) stability of the critical points of the system is investigated, which allows the reconstruction of the history of the Universe in this scenario.
 
\subsection{A short review of dynamical system stability analysis}

In order to obtain an extensive overview of the mathematical analysis of dynamical system techniques, we recommend to the readers the works mentioned in the references \cite{perko2013differential,Wiggins2003}. However, for the sake of self-completeness and self-consistency, we provide here a brief overview of the mathematical methods used in this study. In mathematics, a system of ordinary first-order differential equations (ODEs) is a system consisting of a set of independent variables (functions) depending on one or more dependent variables, related to their derivatives with respect to the independent variables. An autonomous dynamical system is a system of differential equations in which each ODE does not explicitly depend on the independent variable. Usually, the dynamical system is useful to study the dynamics, and to analyze mathematically all the possible ways of a system evolution. 

A dynamical system can be written as a system of ODEs as 
\begin{equation}\label{eq14}
    \dot{X_i}\left(\sigma\right)=F_i \left(X_1\left(\sigma\right),X_2\left(\sigma\right),...,X_n\left(\sigma\right)\right) \hspace{0.35cm},i=1,2,...,n,
\end{equation}
where the dot  represents the derivative with respect to the independent variable $\sigma$, and $F_i \left(i=1,2,..,n\right)$ are $n$ well-defined functions of the dependent variable $X_i, i=1,2,..,n$. Thus, at a particular value of $\sigma$, the state of the system is described by the value of the dependent variable $X_i\left(\sigma\right)$.
The phase space of the dynamical system is the collection of all possible states of the system. Now, for a dynamical system, there may exist some particular states in which the system is in equilibrium. These particular states are known as the critical points, stationary points, or equilibrium points of the system.

Mathematically, the critical points of the dynamical system \eqref{eq14} are the solution of the system of equation, which is generated by equating the r.h.s of \eqref{eq14} with zero. i.e., a point $\left( X^\ast _1\left(\sigma\right), X^{\ast}_2\left(\sigma\right),..., X^{\ast}_n\left(\sigma\right)\right)$ for fixed values of $\sigma$ is called a critical points if the condition, $F_i \left(X^\ast _1\left(\sigma\right), X^{\ast}_2\left(\sigma\right),..., X^{\ast}_n\left(\sigma\right)\right)=0$, for $i=1,2,..,n$ is satisfied. The stability of a dynamical system is analyzed through its critical points.

A critical point $X^\ast_i \left(\sigma\right)$ of the dynamical system \eqref{eq14} is said to be stable if for arbitrary $\epsilon>0 ,\exists \hspace{0.1cm} \delta\left(\epsilon\right)>0$ such that if $Y_i\left(\sigma\right)$ is any solution of \eqref{eq14} satisfying $\lvert \lvert Y_i\left(\sigma_0\right)-X^\ast_i\left(\sigma_0\right)\rvert\rvert<\delta$, then it must also satisfy the condition $\lvert\lvert Y_i\left(\sigma\right)-X^\ast_i\left(\sigma\right)\rvert\rvert <\epsilon \hspace{0.1cm},\hspace{0.1cm} \forall \hspace{0.1cm} \sigma>\sigma_0$.

A critical point $X^\ast_i\left(\sigma\right)$ of the system \eqref{eq14} is considered to be asymptotically stable if for arbitrary $\epsilon>0,\exists \hspace{0.1cm} \delta\left(\epsilon\right)>0$ such that if $Y_i\left(\sigma\right)$ is any solution of the corresponding system satisfying $\lvert\lvert Y_i\left(\sigma_0\right)-X^\ast_i\left(\sigma_0\right)\rvert\rvert<\delta\left(\epsilon\right)$, then it must satisfy $\lim_{\sigma \to \infty}Y_i\left(\sigma\right)=X^\ast_i\left(\sigma\right)$.

In terms of physical interpretation, the key difference between stable and asymptotically stable critical points is that for the stable critical point, the neighborhood trajectories stay close to the critical point over time, while for asymptotically stable critical point not only the neighborhood trajectories stay close to the critical points, they also converge to the critical points as time progresses. A critical point that does not satisfy the condition of stability is called an unstable critical point.

A general technique can be used to determine the stability of the critical points, known as linear stability theory. According to this theory, the stability criteria can be found by calculating the eigenvalues of the Jacobian matrix corresponding to each critical point. The Jacobian matrix for the dynamical system \eqref{eq14} is defined as 
\begin{center} $J
(X_1,X_2,\dots ,X_n)=\begin{pmatrix}
\frac{\partial F_1}{\partial X_1}&\frac{\partial F_1}{\partial X_2}&\dots&\frac{\partial F_1}{\partial X_n} \\
\frac{\partial F_2}{\partial X_1}&\frac{\partial F_2}{\partial X_2}&\dots&\frac{\partial F_2}{\partial X_n}\\
\vdots&\vdots&\vdots&\vdots\\
\frac{\partial F_n}{\partial X_1}&\frac{\partial F_n}{\partial X_2}&\dots&\frac{\partial F_n}{\partial X_n}\\
\end{pmatrix}$\\
\end{center}\par

Let $\lambda^\ast_i$ be the eigenvalues of the Jacobian matrix $J\left(X^\ast_1,X^\ast_2,\dots,X^\ast_n\right)$ for a particular fixed point $X^\ast_i$. Then by the linear stability theory the stability conditions corresponding to $X^\ast_i$ depend on $\lambda^\ast_i$ in the following way
\begin{itemize}
    \item If the real part of all the eigenvalues is positive, i.e. $\Re\left(\lambda^\ast_i\right)>0,\forall i=1,2,\dots,n$, then $X^\ast_i$ is unstable.
    \end{itemize}
    \begin{itemize}
    \item If the real part of all the eigenvalues is negative, i.e. $\Re\left(\lambda^\ast_i\right)<0,\forall i=1,2,\dots,n$, then $X^\ast_i$ is stable.
    \end{itemize}
    \begin{itemize}
    \item If the real part of some of the eigenvalues is positive and the rest are negative, then $X^\ast_i$ is neither stable nor unstable, and it is called a saddle point.
    \end{itemize}

If the set of eigenvalues of the Jacobian matrix corresponding to a critical point consists only of one vanishing eigenvalue, then the particular critical point is normally hyperbolic and its stability will depend on the signature of the rest of non-vanishing eigenvalues \cite{coley2013dynamical}. If more than one eigenvalue vanishes, i.e., $\lambda^\ast_i=\lambda^\ast_j=0,$ for some $i\neq j$, then the critical point is called non-hyperbolic, and linear stability theory fails to determine the stability criteria. For non-hyperbolic critical points. In this case, the central manifold theory or the Lyapunov function are used to determine the stability. However, for a higher-dimensional dynamical system, calculating the corresponding center manifold or finding the appropriate Lyapnouv function is sometimes difficult.
However, in the present manuscript, we study the stability of hyperbolic critical points through the linear stability theory. Due to the non-linear nature of field equations, analytical solutions are sometimes difficult to obtain in cosmology, but we can transform the nonlinear field equations into an autonomous nonlinear dynamical system by employing dynamical variables.\\

In the context of cosmology, the critical points can be associated with several epochs in the cosmological timeline. For an ideal cosmological model, the critical points should represent some of the following cosmological eras: Inflation $\to$ Radiation era $\to$ matter-dominated era $\to$ late time -acceleration era.
The critical points associated with inflation are considered to be unstable, and for matter- and radiation-dominated epochs the corresponding critical points are of saddle nature. The critical point associated with the late-time acceleration is considered to be stable.

\subsection{Dynamical system formulation of the $f\left(R,\mathcal{G},T\right)$ gravity cosmology}

In the present Section we study the detailed dynamics and the corresponding cosmological evolution in the previously introduced $f\left(R,\mathcal{G},T\right)$ gravity, through the dynamical system analysis, which provides a powerful mathematical technique to study the highly nonlinear field equations. It also allows us to study the evolution of many solutions over time, and it can reveal significant information about dark energy, dark matter, and the early Universe.

In order to construct the autonomous dynamical system for the cosmological model of the $f\left(R,\mathcal{G},T\right)$ gravity, we introduce the following  dimensionless variables
\bea\label{var}
    x_1&=&\frac{\dot{f_R}}{H f_R},x_2=\frac{f}{6f_R H^2},x_3=\frac{R}{6H^2},x_4=\frac{\rho}{3H^2 f_R},\nonumber\\
    x_5&=&\frac{\rho f_T}{3H^2 f_R},x_6=\frac{4H^2 f_G}{f_R},x_7=\frac{4H\dot{f_\mathcal{G}}}{f_R}.
\eea

Using the above variable, we can express the first field equation \eqref{eq7} as
\begin{equation}\label{Friedmann constain}
    x_4+x_5+x_3-x_2-x_1+\left(x_3-1\right)x_6-x_7=1.
\end{equation}

The effective energy density corresponding to the matter sector is $\Omega_m=\frac{\rho}{3H^2f_R}$, and the energy density corresponding to the geometric dark energy sector that is related to the effective quantity constructed from the generalized Friedmann equations of the present theory $f\left(R,\mathcal{G},T\right)$ is $\Omega_\Lambda$. Now, the expressions of $\Omega_m$ and $\Omega_\Lambda$, and their relation in terms of the variables (\ref{var}) are obtained from the first Friedmann equation \eqref{eq7} as
\begin{eqnarray}
\Omega_m+\Omega_\Lambda=1,
\eea
where $\Omega_m=x_4$ and $ \Omega_\Lambda=1-x_4$.
From the expression of the Ricci scalar $R$, given in Eq.~\eqref{eq5}, the ratio of the Hubble parameter $H$ and its time derivative can be written in terms of the variable (\ref{var}  as 
\begin{equation}\label{eq19}
    \frac{\dot{H}}{H^2}=\frac{R}{6H^2}-2=x_3-2.
\end{equation}

Using Eq.~\eqref{eq19}, the expressions of the total effective Eos parameter $\omega_{tot}$ and of the deceleration parameter $q$ are found to be
\begin{eqnarray}
    \omega_{tot}&=&-1-\frac{2\dot{H}}{3H^2}=\frac{1}{3}\left(1-2x_3\right),\\
    q&=&-1-\frac{\dot{H}}{H^2}=1-x_3\label{eq dec}.
\end{eqnarray}

Now by differentiating the dynamical variables with respect to $N=\log a\left(t\right)$ and using the conservation equation, we get the autonomous dynamical system equivalent to the generalized Friedmann equations of the considered  $f\left(R,\mathcal{G},T\right)$ gravity model as
\begin{eqnarray}
    \frac{dx_1}{dN}&=&\Gamma-x_1 \left(x_3-2\right)-x_1^2 \label{eq16}\\
    \frac{dx_2}{dN}&=&\Sigma-2\left(x_3 -2\right)x_2 -x_1 x_2\\
    \frac{dx_3}{dN}&=& \mho-2\left(x_3 -2\right)x_3\\
    \frac{dx_4}{dN}&=&-3x_4-x_4 x_1-2x_4\left(x_3 -2\right)\\
    \frac{dx_5}{dN}&=&-\frac{1}{2}x_5-2x_3x_5-x_5x_1\\
    \frac{dx_6}{dN}&=&2\left(x_3 -2\right)x_6 +x_7-x_6 x_1\\
    \frac{dx_7}{dN}&=&\left(x_3-2\right)x_7+4\Xi-x_7 x_1\label{eq23}
\end{eqnarray}

In the dynamical system \eqref{eq16} -\eqref{eq23}, we have defined the following parameters
\begin{equation}
    \Gamma=\frac{\Ddot{f_R}}{H^2 f_R},\mho=\frac{\dot{R}}{6H^3},\Sigma=\frac{\dot{f}}{6f_R H^3},\Xi=\frac{\Ddot{f_\mathcal{G}}}{f_R}.
\end{equation}

Due to the presence of the parameters $\left(\Gamma,\mho,\Sigma,\Xi \right)$, the above dynamical system is not closed. To close it, we have to consider
a specific form of the functional $f\left(R,\mathcal{G},T\right)$. We have chosen a specific form of the function $f$ as $f\left(R,\mathcal{G},T\right)=\alpha_1\mathcal{G}^m+\alpha_2R^{\beta}-2\alpha_3\sqrt{-T} $, where $\alpha_i , (i=1,2,3)$ are the coupling parameters and $\beta$ is the model parameter.
The next step is to find the expression of $\mho$ in terms of $x_i ,\left(i=1,2,..,7\right)$. To determine the expression of $\mho$, we define an additional parameter $\eta$, so that $\eta=\frac{Rf_{,RR}}{f_{,R}}$. Also, the relation between the dynamical variable $x_1$ and $x_3$ can be written as
\begin{eqnarray}\label{eq28}
    x_1 x_3 =\left(\frac{\dot{f_R}}{Hf_R}\right)\left(\frac{R}{6H^2}\right)=\left(\frac{\dot{f_R}}{f_R}\right)\left(\frac{R}{6H^3}\right).
\end{eqnarray}
Using Eq.~\eqref{eq28}, and the definition of $m$, we obtain
\begin{eqnarray}
    \frac{x_1 x_3}{\eta}=\frac{1}{6H^3} \frac{\dot{f_R}}{f_{RR}}=\frac{\dot{R}}{6H^3}=\mho,
    \eea
where
\bea
 \eta=\frac{R f_{RR}}{f_R}=\beta-1.
\end{eqnarray}

Finally, the expression of $\mho$ can be written as 
\begin{equation}\label{eq32}
    \mho=\frac{x_1x_3}{\beta-1}, \beta \neq1.
\end{equation}
Also the specific choice of the function $f\left(R,\mathcal{G},T\right)$ allow us to write the following relation between the dynamical variable as
\begin{eqnarray}
  x_1&=& x_4+x_5+x_3-x_2+\left(x_3-1\right)x_6-x_7-1 \label{extra1} \\
  x_2&=&\frac{x_3}{\beta}+\frac{1}{m}x_6\left(x_3-1\right)-x_5\label{extra2}\\
  x_7&=&\frac{\left(m-1\right)x_6}{\left(x_3-1\right)}\left(\frac{x_1 x_3}{\beta-1}+2\left(x_3-2\right)^2\right)\label{extra3}
\end{eqnarray}
Now by solving the equations \eqref{extra1}, \eqref{extra2}, \eqref{extra3}, we can find the variable $x_1$ and $x_7$ interms of other dynamical variable as
{\scriptsize
\begin{eqnarray}
 x_1&=&\frac{(\beta -1) \left(x_3-1\right) \left(-\frac{\left(x_3-1\right) x_6}{m}-\frac{x_3}{\beta }+x_3+x_4+2 x_5-1\right)}{-\beta +x_3 \left(\beta +(m-1) x_6-1\right)+1}+\nonumber\\
 &&\frac{(\beta -1) \left(x_6 \left(x_3 \left((3-2 m) x_3+8 m-10\right)-8 m+9\right)\right)}{-\beta +x_3 \left(\beta +(m-1) x_6-1\right)+1}\label{extra4}\nonumber\\
 &&\\
 x_7&=&\frac{(m-1) x_6 \left(8 (\beta -1)+x_3 \left(-8 \beta -\frac{x_3}{\beta }+x_4+7\right)\right)}{-\beta +x_3 \left(\beta +(m-1) x_6-1\right)+1}+\nonumber\\
 &&\frac{(m-1) x_6 \left(x_3 \left(-\frac{\left(x_3-1\right) x_6}{m}+x_3 \left(2 \beta +x_6-1\right)+2 x_5-x_6\right)\right)}{-\beta +x_3 \left(\beta +(m-1) x_6-1\right)+1}\nonumber\\
&& \label{extra5}
\end{eqnarray}
}
From equation \eqref{extra2},\eqref{extra4},\eqref{extra5}, it is evident that the variables $x_1,x_2$, and $x_7$ depend on the rest of the variables. Therefore, the differential equation corresponding to $x_1,x_2,x_7$ in the generalized dynamical system \eqref{eq16}-\eqref{eq23} will be reduced and by using \eqref{eq32}, \eqref{extra2}, \eqref{extra4}, \eqref{extra5},we obtain the following four dimensional dynamical system as
\begin{widetext}
{\small
\begin{eqnarray}
 \frac{dx_3}{dN} &=& x_3 \left( 
 \frac{(x_3 - 1) \left( -\frac{(x_3 - 1)x_6}{m} - \frac{x_3}{\beta} + x_3 + x_4 + 2x_5 - 1 \right) + x_6 \left( x_3((3 - 2m)x_3 + 8m - 10) - 8m + 9 \right)}{-\beta + x_3(\beta + (m - 1)x_6 - 1) + 1} 
 - 2x_3 + 4 \right) \label{eq61} \\ 
 \frac{dx_4}{dN} &=& x_4 \Bigg( 
 \frac{(\beta - 1) \left( \beta (m - 1)x_6 \left( x_3((2m - 1)x_3 - 8m + 2) + 8m - 1 \right) 
 - m(x_3 - 1)\left( (\beta - 1)x_3 + \beta(x_4 + 2x_5 - 1) \right) \right)}{\beta m \left( -\beta + x_3(\beta + (m - 1)x_6 - 1) + 1 \right)} \notag \\ 
 &&  - 2x_3 + 1 \Bigg) \\
 \frac{dx_5}{dN} &=& \frac{1}{2} x_5 \Bigg( 
 \frac{2(\beta - 1) \left( \beta (m - 1)x_6 \left( x_3((2m - 1)x_3 - 8m + 2) + 8m - 1 \right) 
 - m(x_3 - 1)\left( (\beta - 1)x_3 + \beta(x_4 + 2x_5 - 1) \right) \right)}{\beta m \left( -\beta + x_3(\beta + (m - 1)x_6 - 1) + 1 \right)} \\
 && - 4x_3 - 1 \Bigg)  \\
 \frac{dx_6}{dN} &=& x_6 \Bigg( 
 \frac{(\beta - 1)(8m - 9)}{-\beta + x_3(\beta + (m - 1)x_6 - 1) + 1} \notag  \quad + \frac{x_3^2(-\beta + 2\beta m + m)\left((\beta - 1)m + \beta(m - 1)x_6\right)}{\beta m \left( -\beta + x_3(\beta + (m - 1)x_6 - 1) + 1 \right)} \notag \\
&&\quad + \frac{x_3 \left( m\left( \beta(8\beta - 8\beta m + 7m - 8) + \beta(x_4 + 2x_5)(m - \beta) + 1 \right) 
 - \beta(m - 1)x_6(-2\beta + (8\beta - 7)m + 1) \right)}{\beta m \left( -\beta + x_3(\beta + (m - 1)x_6 - 1) + 1 \right)} \notag \\
&&\quad + \frac{(\beta - 1)\beta \left( m((8m - 9)x_6 + x_4 + 2x_5) + x_6 \right)}{\beta m \left( -\beta + x_3(\beta + (m - 1)x_6 - 1) + 1 \right)} 
 - 4 \Bigg)
 \label{eq66}
\end{eqnarray}
}
\end{widetext}

\subsection{Critical points and their stability}
The next step in our analysis is to obtain the critical points of the dynamical system mentioned above. To obtain them, we set the right-hand side of Eqs.~\eqref{eq61}-\eqref{eq66} to zero. In this way, we have found a total of eight critical points, denoted as $P_1,P_2,\dots,P_{8}$.

The expressions of the critical points, together with their existence conditions, are presented in Table~\ref{Critical points table}. During the analysis of each critical point, we have associated them with some particular epoch in the cosmological timeline by evaluating their energy densities, as well as other cosmological parameters. In addition, the stability condition of each critical point is individually analyzed using the linear stability theory.

Finally, one can note that these specific choices of the dynamical system variables do not necessarily imply that the effective matter density parameter needs to satisfy the usual standard existence condition, i.e.,  $0\leq\Omega_m=x_4\leq1$. However, because of the complexity of the phase space, we will take into account in our analysis the requirement of the standard existence conditions as a fundamental prerequisite for the validity of the interpretation of the associated critical points.
\begin{table*}
\centering
\begin{tabular}{|c|c|c|c|c|c|}
\hline
&&&&&\\
         Critical point &$x_3$&$x_4$&$x_5$&$x_6$&Existence  \\
			&&&&&Condition\\
			\hline
			\hline
            &&&&&\\
			$P_1$&$0$&$0$&$0$&$0$&Always \\
            &&&&&\\
			\hline
			&&&&&\\
			$P_2$&$0$&$0$&$\frac{1}{4}$&$0$&Always\\
            &&&&&\\
			\hline
            &&&&&\\
            $P_3$&$2$&$0$&$0$&$\frac{2 m-\beta  m}{\beta  (m-1)}$&$m\neq1\land\beta\neq0$\\
            &&&&&\\
            \hline
            &&&&&\\
            $P_4$&$0$&$0$&$0$&$\frac{5 m-8 m^2}{8 m^2-9 m+1}$&$m=\frac{1}{8}\land m=1$\\
            &&&&&\\
            \hline
            &&&&&\\
            $P_5$&$1$&$0$&$0$&$\frac{-2 \beta ^2+2 \beta +1}{2 (\beta -2) \beta }$&$\beta\neq0\land\beta\neq2$\\
            &&&&&\\
            \hline
            &&&&&\\
            $P_6$&$\frac{4 \beta -3}{2 \beta }$&$\frac{-8 \beta ^2+13 \beta -3}{2 \beta ^2}$&$0$&$0$&$\frac{1}{16} \left(13-\sqrt{73}\right)\leq \beta \leq \frac{3}{10}\lor 1\leq \beta \leq \frac{1}{16} \left(\sqrt{73}+13\right)$\\
            &&&&&\\
            \hline
            &&&&&\\
            $P_{7}$&$\frac{8 \beta -9}{4 \beta }$&$0$&$\frac{-22 \beta ^2+35 \beta -9}{8 \beta ^2}$&$0$&$\beta\neq0$\\
            &&&&&\\
\hline
&&&&&\\
$P_8$&$\frac{\beta  (4 \beta -5)}{2 \beta ^2-3 \beta +1}$&$0$&$0$&$0$&$\beta\neq\frac{1}{2}\land\beta\neq1$\\
&&&&&\\
\hline
\end{tabular}
\caption{Critical points along with their existence condition for \( f(R,G,T) = \alpha_1 \mathcal{G}^m + \alpha_2 R^{\beta} - 2\alpha_3 \sqrt{-T} \) gravity }\label{Critical points table}
\end{table*}
\begin{itemize}
    \item \textbf{Critical point $P_1$: }
    The solution corresponding to the critical point $P_1$ always exists in the phase space, since the coordinates are constant and do not depend on other cosmological parameters. The value of the matter density parameter at $P_1$ is $\Omega_m=0$, and the effective Eos parameter is $\omega_{tot}=\frac{1}{3}$. 
    
    Therefore, the critical point corresponds to a radiation-dominated epoch of the Universe. In addition, the deceleration parameter has a constant value $q=1$, which describes a decelerated cosmological expansion. The eigenvalues of the Jacobian matrix at $P_1$ are 
    \begin{equation*}
      \left\lbrace\frac{1}{2},2,5-8 m,\frac{4 \beta -5}{\beta -1}\right\rbrace.  
    \end{equation*}
   Since the set of eigenvalues contains all nonzero eigenvalues, this critical point is hyperbolic. Due to the existence of some positive eigenvalues, this critical point cannot be stable. Since some of the eigenvalues are sensitive to the parameters $m$ and $\beta$, by choosing different combinations of $\left(m,\beta\right)$, this critical point may exhibit a saddle behavior. In Figure \ref {Region plot 1}, we have presented a region in $\left(m,\beta\right)$ space, in which this critical point exhibits a saddle behavior.
\end{itemize}
\begin{figure}
    \centering
    \includegraphics[width=0.78\linewidth]{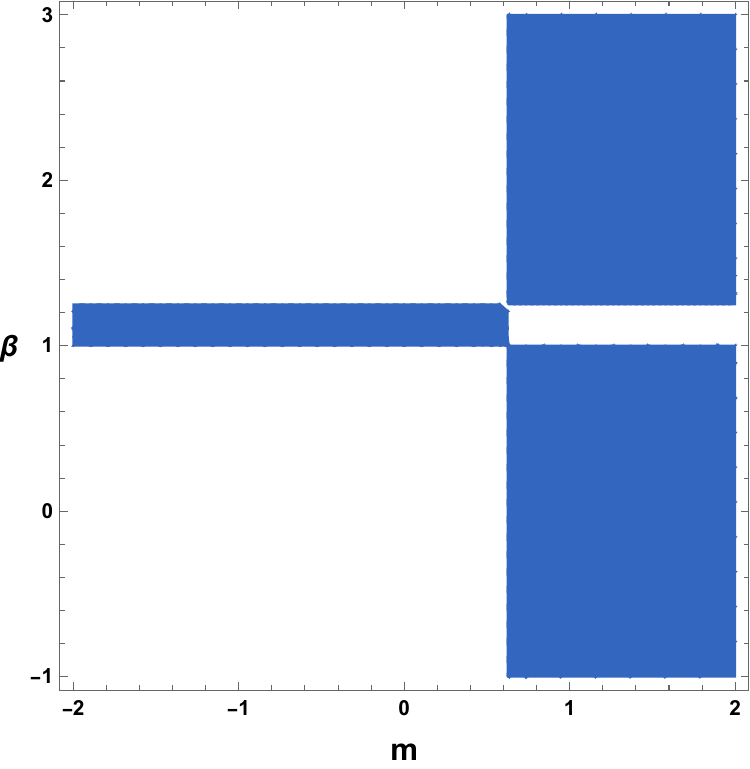}
    \caption{Saddle region for critical point $P_1$}
    \label{Region plot 1}
\end{figure}
    \begin{itemize}
        \item \textbf{Critical point $P_2$:} The cosmological solution corresponding to the second critical point $P_2$ always exists. The value of the matter density $\Omega_m=0$, and this result shows that for this particular solution the matter sector does not contribute to the cosmological dynamics. The values of the effective Eos parameter $\omega_{tot}=\frac{1}{3}$ and of the deceleration parameter $q=1$ indicate that, like in the case of the previous critical point $P_1$, the critical point $P_2$  also describes a radiation-dominated decelerated era of the Universe. The set of eigenvalues of the Jacobian matrix corresponding to $P_2$ is
        \begin{equation*}
          \left\lbrace-\frac{1}{2},\frac{3}{2},\frac{1}{2} (9-16 m),\frac{8 \beta -9}{2 (\beta -1)}\right\rbrace.   
        \end{equation*}
        
        Due to the presence of both negative and positive eigenvalues, the stability of this solution is not possible and it will always exhibit saddle behavior.
    \end{itemize}

    \begin{itemize}
        \item \textbf{Critical point $P_3$:} The cosmological solution corresponding to the critical point $P_3$ describes the effects of the Ricci scalar curvature $R$ and the Gauss-Bonnet invariant $\mathcal{G}$ on the dynamics. The value of the matter density parameter $\Omega_m=0$, and, consequently, $\Omega_\Lambda=1$. Hence,  this particular solution describes a completely geometric dark energy-dominated Universe, which is usually related to the late-time acceleration era. The values of the total Eos parameter and the deceleration parameters are $\omega_{tot}=-1$ and $q=-1$, respectively. Hence, for this solution, the geometric dark energy component mimics the behavior of the cosmological constant. This solution can describe the current cosmological state of the Universe.  The negative value of the deceleration parameter indicates that the present expansion of the Universe is accelerating. To determine the stability conditions, the eigenvalues of the Jacobian matrix corresponding to critical point $P_3$ are obtained as
\begin{align*}
\Bigg\{ 
&  \frac{3 \left(-\beta ^2+\beta +2 \beta  m-4 m\right)}{\beta ^2-\beta -2 \beta  m+4 m}, \\
&\frac{9 \left(-\beta ^2+\beta +2 \beta  m-4 m\right)}{2 \left(\beta ^2-\beta -2 \beta  m+4 m\right)}, E_3,E_4
\Bigg\}
\end{align*} 
   Due to the large mathematical expression of $m$ and $\beta$, we do not mention the expression of the third and fourth eigenvalues, namely $E_3$ and $E_4$, explicitly in the manuscript. Since all the eigenvalues are highly sensitive to the parameters $m$ and $\beta$, the stability behavior depends on the value combination of the parameters $m$ and $\beta$. In the Figure \ref{region plot 2}, we have numerically evaluated a region in $\left(m,\beta\right)$ space in which the critical point $P_3$ will exhibit a stable behavior. Consequently, outside these shaded regions, this critical point will exhibit either a saddle or an unstable nature.   
    \end{itemize}
    \begin{figure}[htb]
    \centering
    \includegraphics[width=0.78\linewidth]{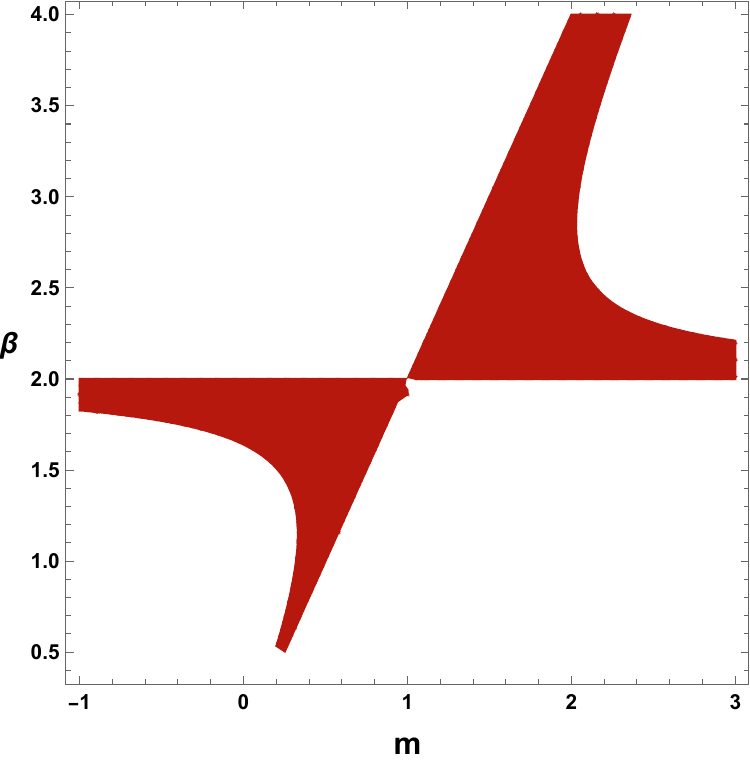}
    \caption{Stable region for critical point $P_3$}
    \label{region plot 2}
\end{figure}
 \begin{itemize}
        \item \textbf{Critical point $P_4$:} The cosmological solution corresponding to $P_4$ describes a decelerated radion-dominated epoch, since $\omega_{tot}=\frac{1}{3}$ and $q=1$. 
     The physical properties corresponding to this critical point are quite similar to the previous critical point $P_1$ and $P_2$.
 To determine the stability condition, the eigenvalues corresponding to the Jacobian matrix are obtained as\\ 
        \begin{equation*}
            \left\lbrace -\frac{4 (2 m-\beta )}{\beta -1},8 m-5,8 m-3,\frac{1}{2} (16 m-9)\right\rbrace,
        \end{equation*}

Since all the eigenvalues of the Jacobian matrix are nonzero,$P_4$ is a hyperbolic critical point. As the expressions of the eigenvalues depend on the parameters $m$ and $\beta$, considering different combinations of parameters $\left(m,\beta\right)$, this critical point can exhibit different stability behavior. In Figure \ref{regionplot 3}, the numerically evaluated stable region in $\left(m,\beta\right)$ is presented. Outside the shaded region in Figure\ref{regionplot 3}, this critical point will exhibit either a saddle or an unstable characteristic.
\end{itemize}
\begin{figure}
    \centering
    \includegraphics[width=0.78\linewidth]{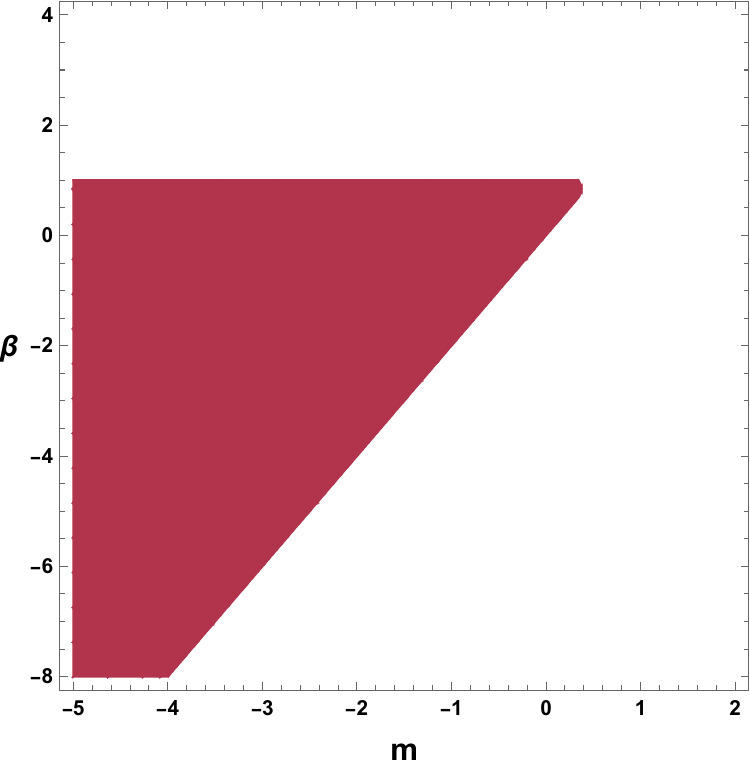}
    \caption{Stable region for critical point $P_4$}
    \label{regionplot 3}
\end{figure}
\begin{itemize}
\item \textbf{Critical point $P_5$:} The cosmological solution corresponding to this critical point represents a cosmological era,in which $\omega_{tot}=-\frac{1}{3}$ and $q=0$. These values of the effective Eos parameter and the deceleration parameter indicate the transitional epoch from the decelerated to the accelerated expansion era.  The eigenvalues of the Jacobian matrix are 
\begin{equation*}
 \left\lbrace\frac{2 (2 m-\beta )}{m-1},2 (\beta -2),2 \beta -3,\frac{1}{2} (4 \beta -9)\right\rbrace
\end{equation*}
The stability features of this critical point depend on the parameters $m$ and $\beta$. A numerically evaluated stable region for the critical point $P_5$ is presented in Figure \ref{region plot 4}. 
\end{itemize}
\begin{figure}
    \centering
    \includegraphics[width=0.78\linewidth]{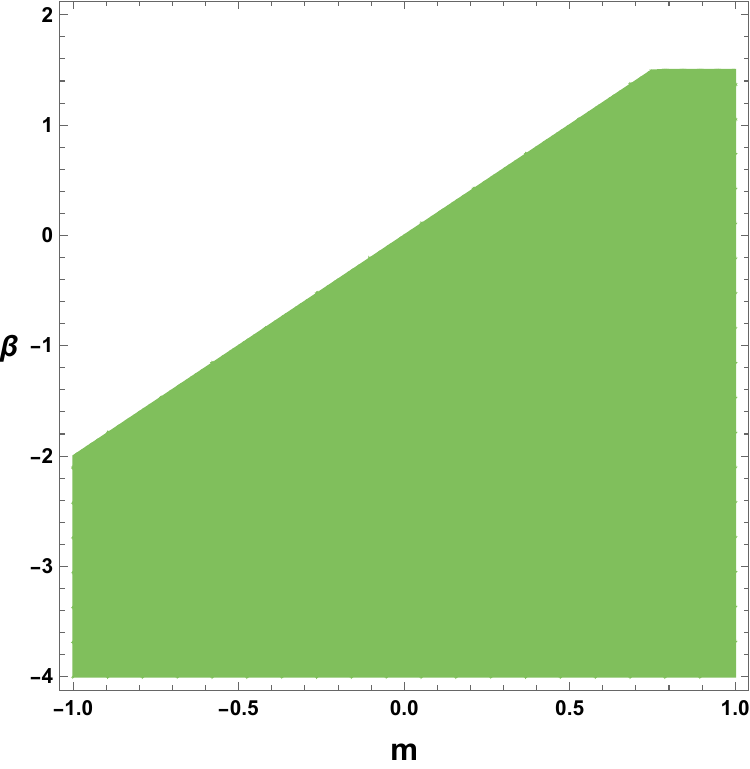}
    \caption{Stable region for critical point $P_5$}
    \label{region plot 4}
\end{figure}
\begin{itemize}
\item \textbf{Critical point $P_6$:} The existence of the critical point $P_6$ depends on the parameter $\beta$. According to the standard existence condition $0\leq\Omega_m\leq1$ .i.e., the effective matter density must lie in the range $[0,1]$. This condition  gives the following constraint on the parameter $\beta$ 
\begin{align*}
&& \frac{1}{16} \left(13-\sqrt{73}\right)\leq \beta \leq \frac{3}{10}\nonumber\\
&& \lor \;1\leq \beta \leq \frac{1}{16} \left(\sqrt{73}+13\right)
\end{align*}
The value of the matter density parameter is 
$\Omega_m=\frac{-8 \beta ^2+13 \beta -3}{2 \beta ^2}$. 
For $\beta=\frac{3}{10}$, we get $\Omega_m=1$, representing a cosmological solution completely dominated by the matter sector. The value of the effective Eos parameter is $\omega_{tot}=\frac{1}{ \beta }-1$. Since the expression of the Eos parameter includes $\beta$, different choices of $\beta$ will lead to different cosmological epochs. For $\beta>\frac{3}{2}\Rightarrow -1<\omega_{tot}<-\frac{1}{3}$, the critical point $P_6$ represents the quintessence era, while for $\beta <0\Rightarrow \omega_{tot}<-1$, it represents the super acclerating phantom era. In addition, this solution can represent an accelerated universe, with $q<0$ for $\beta<0\lor \beta>\frac{3}{2}$. In terms of studying the stability features, the eigenvalues of the Jacobian matrix corresponding to $P_6$ are given by
\begin{widetext}
\begin{align*}
\Bigg\{ 
& -\frac{3 (2 m-\beta )}{\beta },-\frac{3}{2}, \\
&\frac{\pm\sqrt{256 \beta ^8 m^2-864 \beta ^7 m^2+1025 \beta ^6 m^2-498 \beta ^5 m^2+81 \beta ^4 m^2}-3 \beta ^3 m+3 \beta ^2 m}{4 (\beta -1) \beta ^3 m}
\Bigg\}
\end{align*} 
\end{widetext}

All the eigenvalues of the Jacobian matrix are nonzero and complex mathematical expressions of the parameters $m$ and $\beta$. In Figure \ref{regionplot p6}, a numerically evaluated stable region corresponding to this critical point is presented. In the shaded region of Figure \ref{regionplot p6}, all the eigenvalues consist of negative values; therefore, by the linear stability theory, this critical point will be stable inside this shaded region. However, outside this region the critical point will behave as an saddle point or as an unstable node.
\end{itemize}
\begin{figure}
    \centering
    \includegraphics[width=0.78\linewidth]{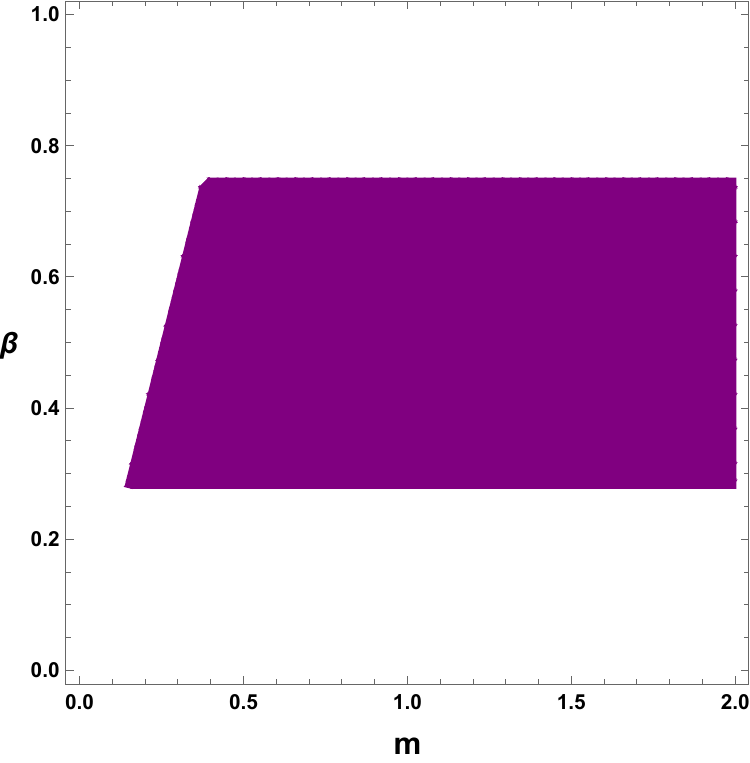}
    \caption{Stable region for critical point $P_6$}
    \label{regionplot p6}
\end{figure}
\begin{itemize}
\item \textbf{Critical point $P_7$:} The critical point $P_7$ corresponds to a cosmological solution that is completely dominated by the geometric dark energy component, as $\Omega_\Lambda=1$. Since $\Omega_m=0$, the matter sector does not contribute to the cosmological dynamics of this particular solution. The total Eos parameter is obtained as a function of the parameter $\beta$ as $\omega_{tot}=\frac{3}{2 \beta}-1$.
    
If we consider $\beta=\frac{3}{2}$, then the value of the Eos parameter vanishes and the corresponding solution will represent the matter-dominated era. On the other hand, considering $\beta>\frac{9}{4}$, we find that the value of the Eos parameter lies in the range $-1<\omega_{tot}<-\frac{1}{3}$ and the corresponding solution will represent the era of quintessence. Furthermore, one can obtain $\omega_{tot}<-1$, called the phantom era, from this critical point by considering $\beta<0$. For an accelerating solution with $q<0$, the value of the parameter $\beta$ must satisfy the condition $\beta<0 \vee \beta>\frac{9}{4}$. The eigenvalues of the Jacobian matrix corresponding to the critical point $P_7$ are
\begin{widetext}
\begin{align*}
\Bigg\{-\frac{9 (2 m-\beta )}{2 \beta },\frac{3}{2},\frac{-\sqrt{1444 \beta ^8 m^2-5412 \beta ^7 m^2+7253 \beta ^6 m^2-4014 \beta ^5 m^2+729 \beta ^4 m^2}+6 \beta ^4 m-15 \beta ^3 m+9 \beta ^2 m}{8 (\beta -1) \beta ^3 m},\nonumber\\ 
  \frac{\sqrt{1444 \beta ^8 m^2-5412 \beta ^7 m^2+7253 \beta ^6 m^2-4014 \beta ^5 m^2+729 \beta ^4 m^2}+6 \beta ^4 m-15 \beta ^3 m+9 \beta ^2 m}{8 (\beta -1) \beta ^3 m} \Bigg\}.
\end{align*}
\end{widetext}

Due to the presence of the positive eigenvalues $E_2=\frac{3}{2}$, stability is not possible for this critical point. Therefore, it will be either a
saddle point or a purely unstable critical point, depending on the value of the parameter $m$ and $\beta$.
We have numerically analyzed the eigenvalues corresponding to the critical point $P_7$ for all possible values of the parameter $m$ and $\beta$ and we have found that some of the eigenvalues are always negative regardless of the value of the parameter $m$ and $\beta$. Hence, according to linear stability theory, the critical point $P_6$ will always exhibit saddle behavior.
\end{itemize}
\begin{itemize}
\item \textbf{Critical point $P_{8}$:} The last critical point $P_8$ represents a cosmological solution that is completely governed by the geometric dark energy component such as $\Omega_m=0$ and $\Omega_\Lambda=1$. The value of the effective Eos parameter is 
\[
\omega_{tot}=\frac{-6 \beta^2+7 \beta +1}{6 \beta^2-9 \beta+3}. 
\]
This critical point is interesting in terms of cosmological evolution, since it can represent almost all possible cosmological eras for different choices of the parameter $\beta$. For 
\[
\beta=\frac{1}{12} \left(7-\sqrt{73}\right)\lor \beta =\frac{1}{12} \left(\sqrt{73}+7\right), 
\]
the solution represents the matter era, and the corresponding Eos parameter is $\omega_{tot}=0$. Consequently, in the range
\[
\beta <\frac{1}{2} \left(1-\sqrt{3}\right)\lor \frac{1}{2} \left(\sqrt{3}+1\right)<\beta<2, 
\]
the solution will represent a quintessence era, while for $\beta =2$ one can recover the de Sitter solution with $\omega_{tot}=-1$, with the geometric dark energy component mimicking the cosmological constant behavior. Moreover, for $\frac{1}{2}<\beta<1\lor \beta >2$, this solution will exhibit a super-accelerating phantom era with $\omega_{tot}<-1$. In order to achieve an accelerated solution with $q<0$, the value of $\beta$ must lie in the following region 
\[
\hspace{0.8cm}\beta<\frac{1}{2} \left(1-\sqrt{3}\right)\lor \frac{1}{2}<\beta<1\lor \beta>\frac{1}{2} \left(\sqrt{3}+1\right). 
\]

Now, for studying  the stability features, the eigenvalues of the Jacobian matrix corresponding to $P_8$ are obtained as
\begin{align*}
\Bigg\{ 
&  -\frac{4 \beta -5}{\beta -1},\frac{2 \left(-\beta ^2+2 \beta +2 \beta  m-4 m\right)}{(\beta -1) (2 \beta -1)}, \\
&-\frac{8 \beta ^2-13 \beta +3}{(\beta -1) (2 \beta -1)},-\frac{22 \beta ^2-35 \beta +9}{2 (\beta -1) (2 \beta -1)}
\Bigg\}
\end{align*} 

All the eigenvalues are non-vanishing and functions of $m$ and $\beta$ only, implying that this critical point is hyperbolic. In Figure \ref{regionplot 8}, the numerically evaluated a region in $\left(m,\beta\right)$ space where all the eigenvalues are negative and hence by linear stability theory in this region the critical point $P_8$ always exhibits stable behavior. Outside this shaded region, this critical point either behaves as a saddle point or as an unstable node. 
\end{itemize}
\begin{figure}
    \centering
    \includegraphics[width=0.78\linewidth]{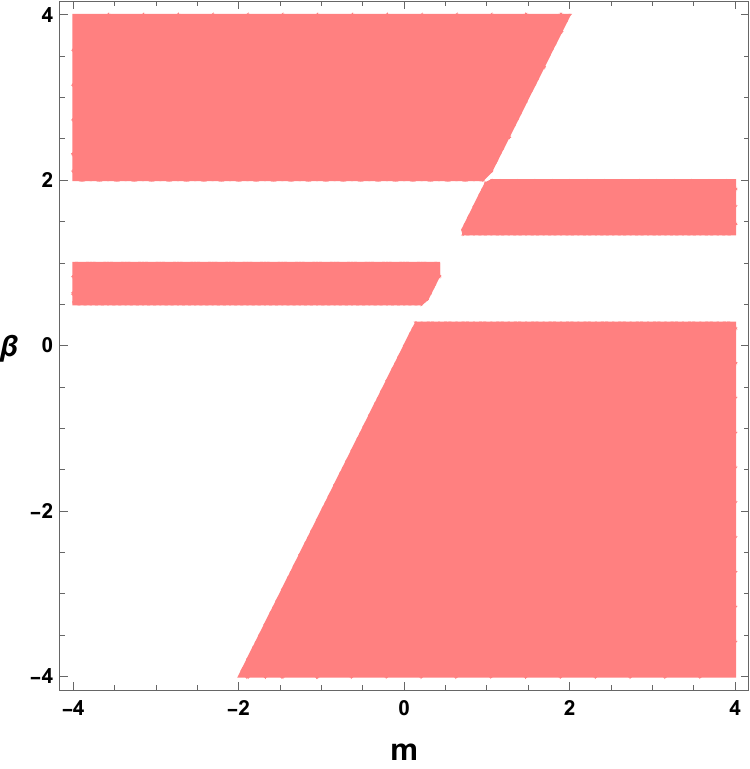}
    \caption{Stable region for critical point $P_8$}
    \label{regionplot 8}
\end{figure}
\subsubsection{Phase space portraits of the critical points}

In Fig.~\ref{phase potrait}, we have presented the diagrams of the phase space projection on dynamical variable, corresponding to each critical point of the dynamical system. As one can see in Figs.~\ref{phase potrait}, the trajectories in the neighborhood of the critical points $P_3,P_4,P_5,P_6,P_8$ are attracted towards these critical points, this feature corresponding to their stable behavior.
Also, one can see that some of the trajectories in the neighborhood of the critical points $P_1,P_2,P_7$ are attracted towards them, while some trajectories are repelled from them. These properties describe the saddle or unstable behavior of the corresponding critical points.
\begin{figure*}[htb]
\begin{subfigure}{.32\textwidth}
\includegraphics[width=\linewidth]{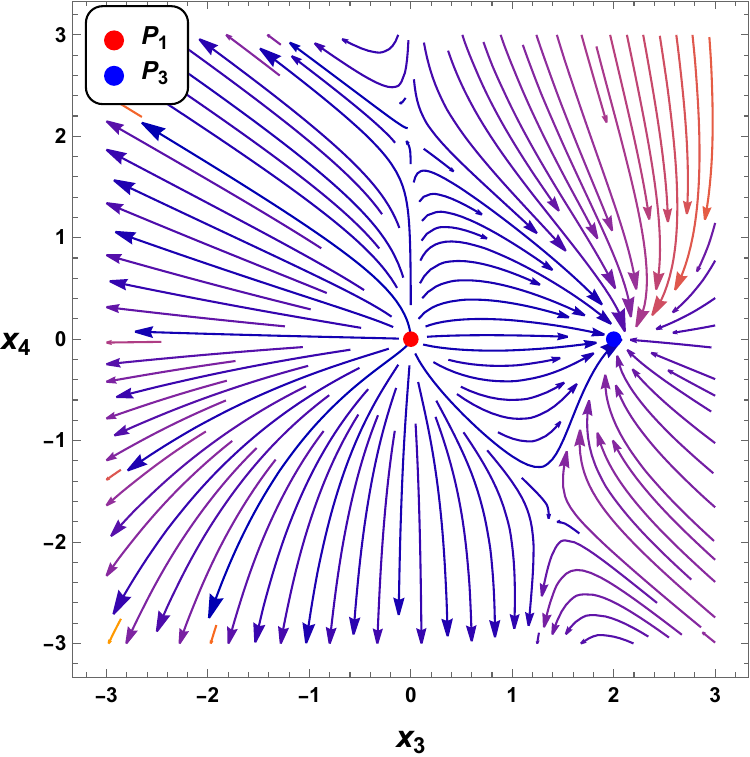}
    \caption{Projection of phase space on the  $x_3-x_4$ plane}
    \label{phase portrait 1}
\end{subfigure}
\hfil
\begin{subfigure}{.32\textwidth}
\includegraphics[width=\linewidth]{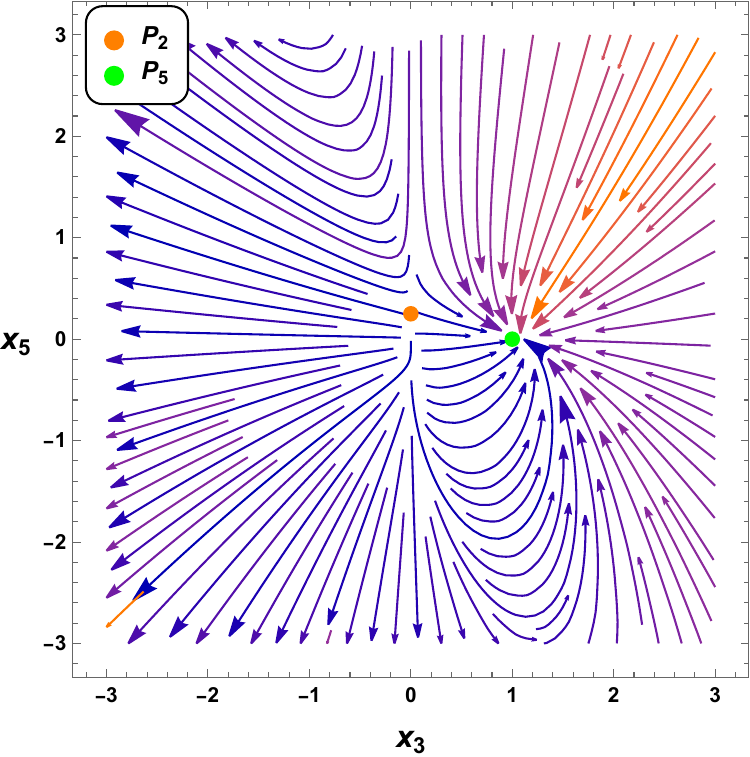}
    \caption{Projection of phase space on the $x_3-x_5$ plane}
    \label{phase portrait 2}
\end{subfigure}
\hfil
\begin{subfigure}{.32\textwidth}
\includegraphics[width=\linewidth]{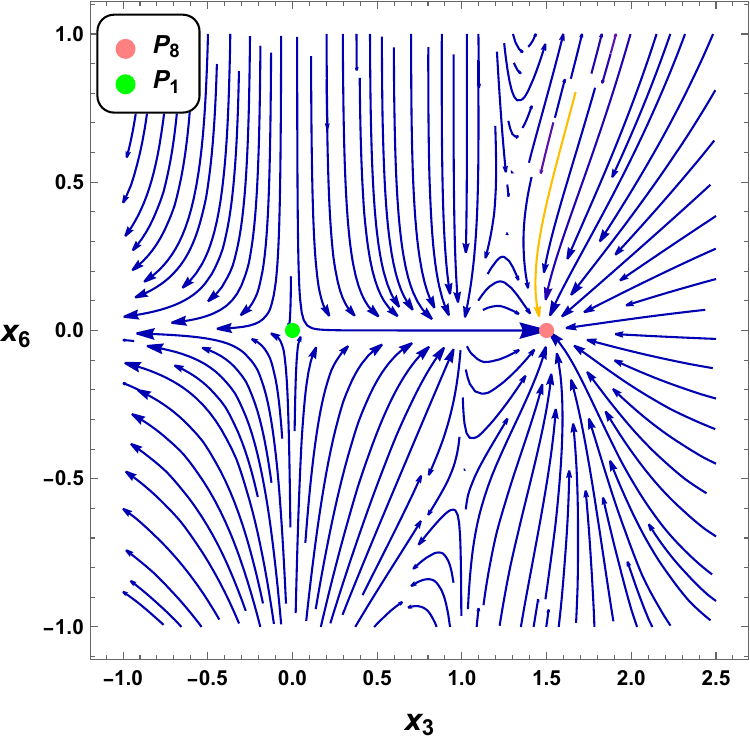}
    \caption{Projection of phase space on the $x_3-x_6$ plane}
    \label{phase portrait 3}
\end{subfigure}
\begin{subfigure}{.32\textwidth}
\vspace{1cm}
\includegraphics[width=\linewidth]{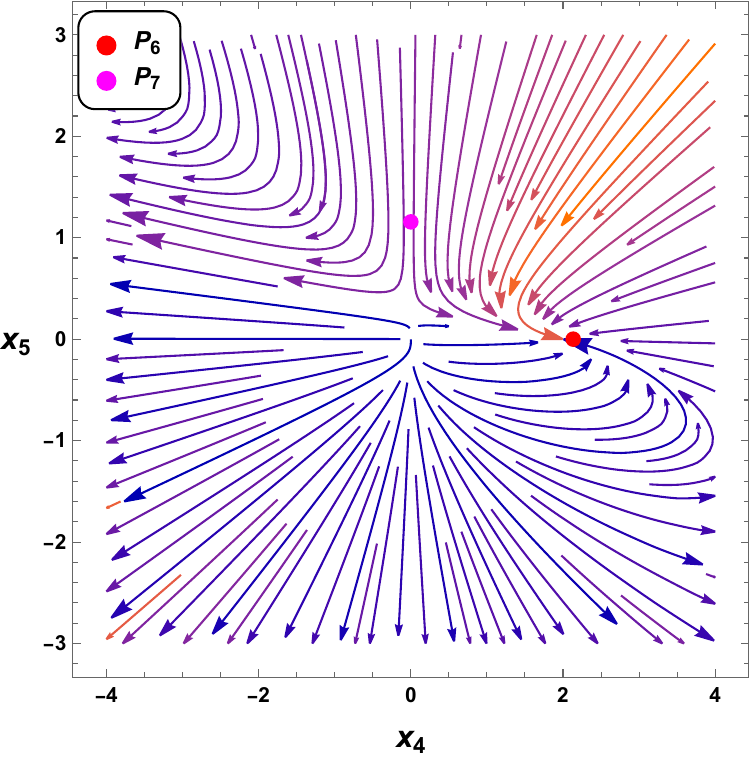}
    \caption{Projection of phase space on the $x_4-x_5$ plane}
    \label{phase portrait 4}
\end{subfigure}
\hfil
\begin{subfigure}{.32\textwidth}
\includegraphics[width=\linewidth]{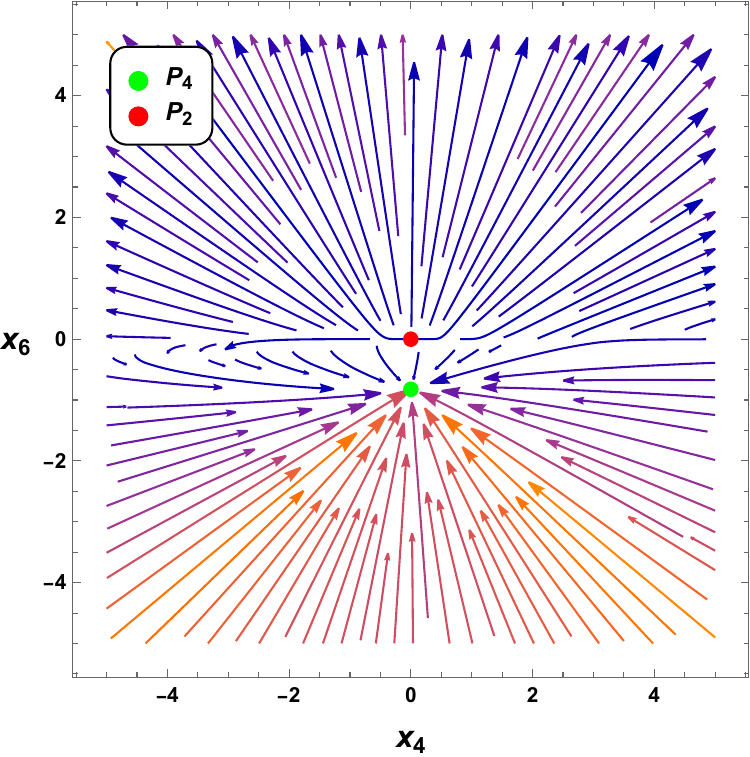}
    \caption{Projection of phase space on the $x_4-x_6$ plane}
    \label{phase portrait 5}
\end{subfigure}
\begin{subfigure}{.32\textwidth}
\includegraphics[width=\linewidth]{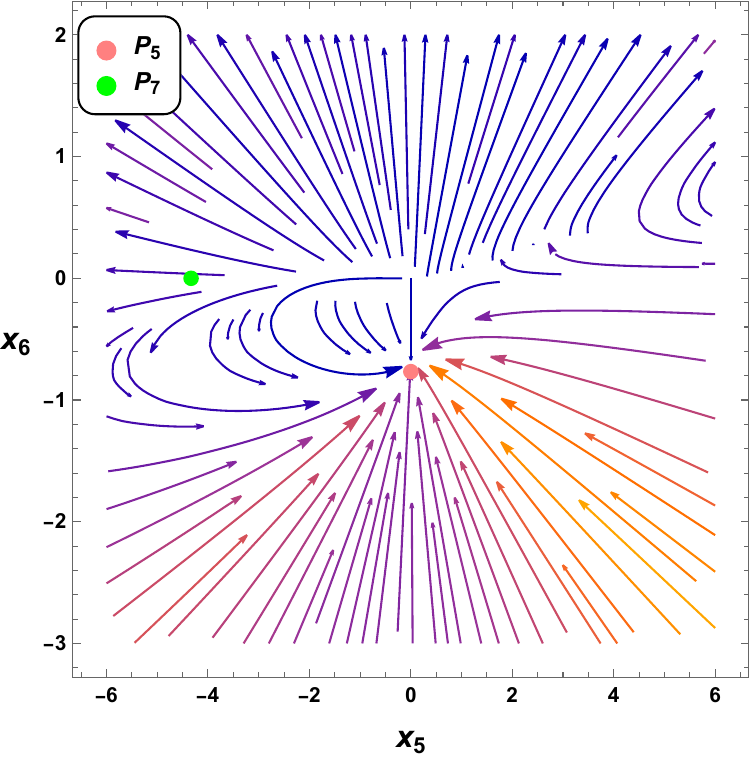}
    \caption{Projection of phase space on the $x_5-x_6$ plane}
    \label{phase portrait 6}
\end{subfigure}
\caption{Projection of phase space diagram corresponding to the critical points $P_i , \left(i=1,2,\dots,8\right)$ }\label{phase potrait}
\end{figure*}
\subsection{Discussion of the cosmological behavior}  
Now, after studying the stability features and dynamical epochs connected with each critical point, it is important to investigate the long-term dynamics of the specific cosmological model. In order to understand the complete picture of dynamics, the evolution of background cosmological parameters like the density parameters $\Omega_m$, $\Omega_\Lambda$, Eos parameter $\omega_{tot}$ and the deceleration parameter $q$ plays a significant role. 

Some recent observational data suggest that our Universe is nearly flat, and the current values of density parameters are approximately $\Omega_{\Lambda 0}\approx 0.7$ and $\Omega_{m0} \approx 0.3$ respectively. In the dynamical system defined by the equations \eqref{eq61}-\eqref{eq66}, numerical integration can be performed with appropriate initial conditions to accurately capture the full cosmological evolution across various epochs. 

We have numerically solved the dynamical system. The numerical solutions describing the density parameters and the Eos parameter together with the deceleration parameter are presented in Fig.~ \ref{fig5}. In Fig.~\ref{fig5}, the vertical line $N=0$ represents the current timeline, and $N>0$ and $N<0$ represent the future and past epochs, respectively. 

The numerical solution corresponding to the density parameter, as shown in Fig.~\ref{fig density}, shows that for the specific $f\left(R,\mathcal{G},T\right)$ gravity model, the current value of matter density is $\Omega_{m0}\approx0.24$, and the geometrical dark energy density $\Omega_{\Lambda0}\approx0.76$, which aligns with the current observational results. 

A key characteristic of the energy density is that in the early universe, matter density dominated dark energy density. However, as the universe evolves, matter density gradually decreases, while dark energy density becomes increasingly dominant. In the far future, dark energy density is projected to completely surpass matter density, i.e., $\Omega_m\to 0$ and $\Omega_\Lambda \to 1$, leading to a Universe entirely dominated by dark energy. 

On the other hand, from Fig.~\ref{fig eos}, it is evident that the evolution of the total equation of state (EoS) parameter $\omega_{tot}$ begins with a pre-matter-dominated era, where $\omega_{tot}>0$. Then it successively transitions through the matter era $\omega_{tot}=0$ to the quintessence era $\left(-1<\omega_{tot}<-\frac{1}{3}\right)$ and the de Sitter era $\left(\omega_{tot}=-1\right)$, respectively. After passing through the de Sitter era,  it finally reaches the phantom era corresponding to $\omega_{tot}<-1$ in the future. 

Thus, the evolution of $\omega_{tot}$ successfully illustrates a complete picture of the cosmological evolution from the pre-matter-dominated era to beyond the de Sitter phase by attaining each important cosmological era one by one. Furthermore,the evolution of the deceleration parameter $q$, as shown in Fig.~\ref{fig eos} reveals a smooth transition from a decelerated expansion phase $\left(q>0\right)$ to an accelerated expansion phase $\left(q<0\right)$ at $N=-0.29$. The present value of the deceleration parameter is found to be approximately $q\approx-0.57$, which is in agreement with observational constraints.
\begin{figure*}[htb]
\begin{subfigure}{.496\textwidth}
\includegraphics[width=\linewidth]{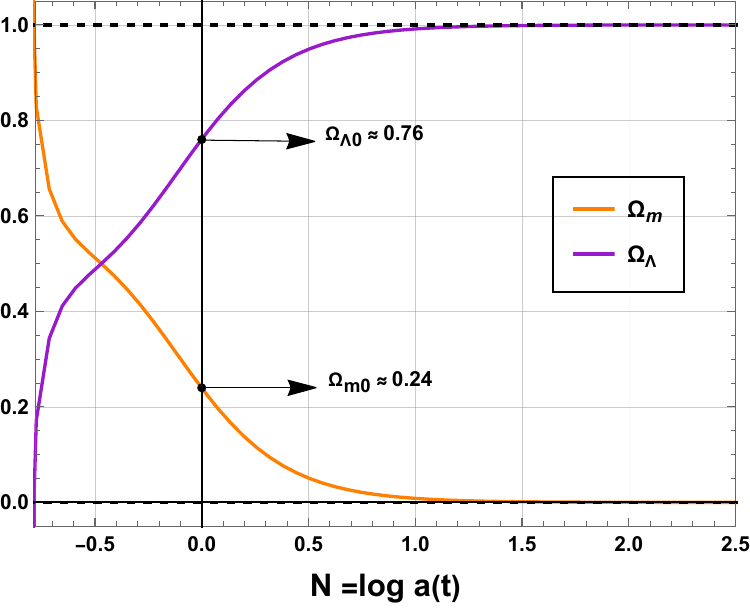}
    \caption{Evolution of density parameters $\Omega_m$ and $\Omega_\Lambda$}
    \label{fig density}
\end{subfigure}
\hfil
\begin{subfigure}{.496\textwidth}
\includegraphics[width=\linewidth]{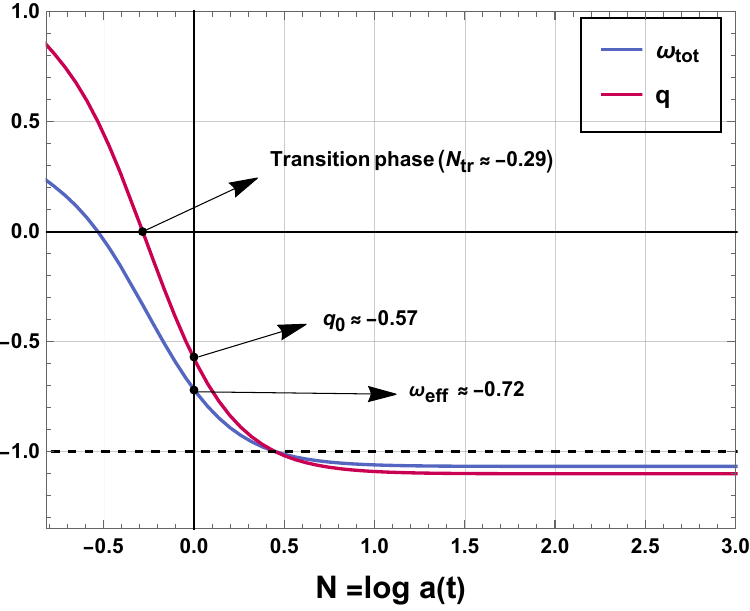}
    \caption{Evolution of Eos parameter $\omega_{tot}$ and of the deceleration parameter $q$}
    \label{fig eos}
\end{subfigure}
\caption{Evolution of various cosmological parameters for $f\left(R,\mathcal{G},T\right)$ gravity model}\label{fig5}
\end{figure*}
\section{Statefinder parameter analysis}\label{sect4}

In this Section, a brief analysis of the statefinder parameters is outlined for the present cosmological model through the dynamical variable formulation. The statefinder parameters, namely the jerk parameter $j$ and the snap parameter or the jounce parameter $s$, are important tools that allow us to unfold different dynamical characteristics of any cosmological model. These parameters are dimensionless and can be constructed from the particular combination of the scale factor and its time derivatives. 

In the hierarchy of geometrical cosmological parameters, the jerk parameter $j$ follows the deceleration parameter $q$ and the Hubble parameter $H$, whereas the snap parameter $s$ is formed by the linear combination of $j$ and $q$ such that it does not rely on the dark energy density.

\subsection{Jerk parameter}

 The jerk parameter appear in the fourth term of the Taylor series of cosmic scale factor $a\left(t\right)$ around particular time $t_0$ as 
\bea
    \frac{a\left(t\right)}{a_0}&=&1+H_0\left(t-t_0\right)-\frac{1}{2}q_0 H_0^2\left(t-t_0\right)^2+\nonumber\\
    &&\frac{1}{6} j_0 H_0^3\left(t-t_0\right)^3 +O[\left(t-t_0\right)^4].
\eea

In light of the existence of numerous dark energy theories, the jerk parameter is a crucial component in the search for an appropriate candidate for the scientific interpretation of cosmic dynamics. The jerk parameter also plays an important role in order to distinguish between various dark energy models and the transition between them. For example, $j=1$ can represent the flat $\Lambda$CDM model of dark energy, and we can obtain the present accelerated solution for $j>0$ with negative values of the deceleration parameter $q$.

Alternatively, the jerk parameter can also be written as a third-order time derivative of the scale factor such that
\begin{eqnarray}
    j=\frac{\dddot{a}}{aH^3}=2q^2+q-\frac{\dot{q}}{H}
\end{eqnarray}

By using some algebraic calculations on the expression of the deceleration parameter $q$ and the dynamical system \eqref{eq16} - \eqref{eq23}, we have obtained the expression of the jerk parameter as a function of the dynamical variables such that
\begin{eqnarray}\label{eq33}
    &&j=3+\frac{x_3 \left(m \left(x_3-1\right) \left((\beta -1) x_3+\beta  \left(-\beta +x_4+2 x_5\right)\right)\right)}{\beta  m \left(-\beta +x_3 \left(\beta +(m-1) x_6-1\right)+1\right)}\nonumber\\
    &&-\frac{x_3 \left(\beta  (m-1) x_6 \left(x_3 \left((2 m-1) x_3-7 m+2\right)+8 m-1\right)\right)}{\beta  m \left(-\beta +x_3 \left(\beta +(m-1) x_6-1\right)+1\right)}\nonumber\\
    &&
\end{eqnarray}
From some recent cosmological observations, the current value of the jerk parameter is found to be very close to 1. The evolution of the jerk parameter for our current $f\left(R,\mathcal{G},T\right)$ gravity is presented in Fig.~\ref{fig7}. As it is evident from Figure\ref{jerk diagram}, the current value of the jerk parameter is obtained as $j_0 \approx 1.05$, which is similar to the value predicted by the $\Lambda$CDM model and satisfies the observational result.
\subsection{Snap parameter}
The snap parameter arises from the fifth term in the Taylor series of scale factor $a(t)$ around a given time $t_0$ as
\begin{eqnarray}
 \frac{a\left(t\right)}{a_0}&=&1+H_0\left(t-t_0\right)-\frac{1}{2}q_0 H_0^2\left(t-t_0\right)^2+\nonumber\\
   && \frac{1}{6} j_0 H_0^3\left(t-t_0\right)^3 +\frac{1}{24}s_0 H_0^4\left(t-t_0\right)^4+O[\left(t-t_0\right)^5]. \nonumber\\ 
\end{eqnarray}
The snap parameter has an important role in understanding the transition between different dark energy models. More specifically, it represents the deviation of any cosmological model from the conventional $\Lambda$CDM models of dark energy. 
In terms of the scale factor $a\left(t\right)$ and its fourth-order time derivative and the deceleration parameter, the snap parameter can be written as
\begin{eqnarray}\label{eq35}
    s=\frac{\ddddot{a}}{a H^4}=\frac{j-1}{3\left(q-\frac{1}{2}\right)}
\end{eqnarray}
Using Eq.~\eqref{eq dec}, and Eqs.~\eqref{eq33} and \eqref{eq35}, we obtain the expression of $s$ in terms of the dynamical variables as
{\small
\begin{eqnarray}
&& s=\frac{2 \beta  (m-1) x_3 x_6 \left(x_3 \left((2 m-1) x_3-7 m+2\right)+6 m-1\right)}{3 \beta  m \left(2 x_3-1\right) \left(-\beta +x_3 \left(\beta +(m-1) x_6-1\right)+1\right)}+\nonumber\\
 &&\frac{2 m \left(x_3-1\right) \left(-2 (\beta -1) \beta -\left((\beta -1) x_3^2\right)+\beta  x_3 \left(\beta -x_4-2 x_5\right)\right)}{3 \beta  m \left(2 x_3-1\right) \left(-\beta +x_3 \left(\beta +(m-1) x_6-1\right)+1\right)}\nonumber\\
 &&
\end{eqnarray}
}
The evolution of the snap parameter $S$, corresponding to our specific $f\left(R,\mathcal{G},T\right)$ model, is presented in Fig.~\ref{fig7}. From the cosmological observation, the current value of the snap parameter is found in the range $-1<s<0$, and for the $\Lambda$CDM model, the value will be $s=0$. We have presented the evolution of the Snap parameter corresponding to the $f\left(R,\mathcal{G},T\right)$ gravity model in Fig.~\ref{snap diagram}. At the present epoch, the value of $s$ for our cosmological model is obtained as $s_0\approx0$, which is in the valid range and close to the observational result. It converges to $0$ as $N\to \infty$, indicating the similarity of the late-time acceleration phase with the $\Lambda$CDM model. 
\begin{figure*}[htb]
\begin{subfigure}{.45\textwidth}
\includegraphics[width=\linewidth]{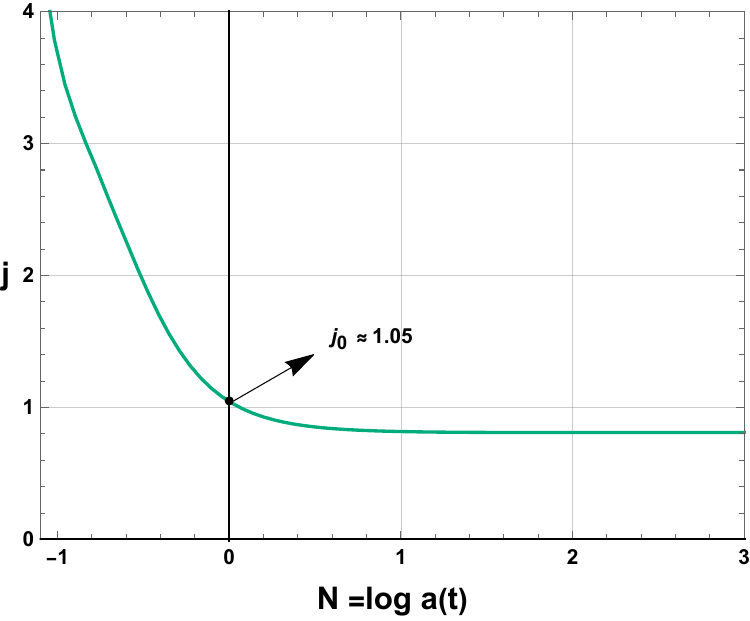}
    \caption{Plots of the jerk parameter $j$ vs $N$}
    \label{jerk diagram}
\end{subfigure}
\hfil
\begin{subfigure}{.46\textwidth}
\includegraphics[width=\linewidth]{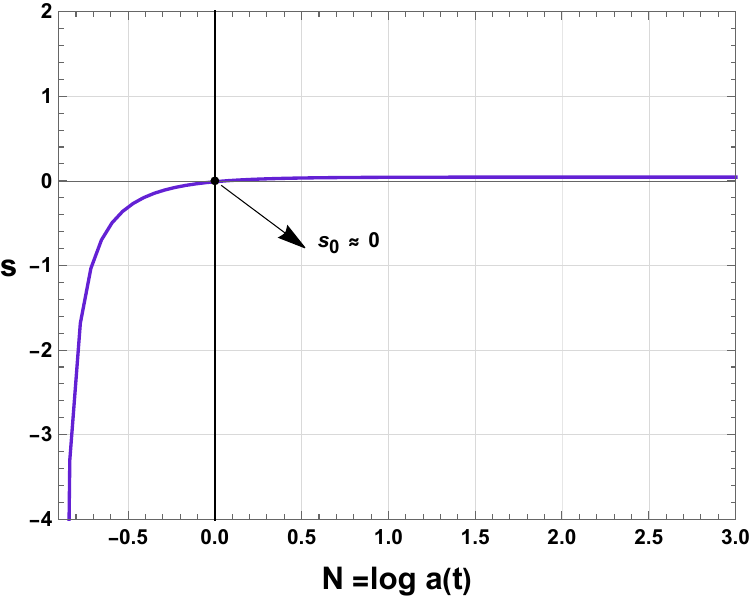}
    \caption{Plots of the snap parameter $s$ vs $N$}
    \label{snap diagram}
\end{subfigure}
\caption{Evolution of various cosmography parameters for $f\left(R,\mathcal{G},T\right)$ gravity model}\label{fig7}
\end{figure*}  
\section{Conclusions}\label{sect5}

The incorporation of the curvature corrections and of the GB invariant, together with the coupling of matter and geometry, may offer some new perspectives on the evolution and dynamics of the early Universe and of its transition to a late-time acceleration. 

By including higher-order curvature corrections, as well as matter-coupling effects, the GB modified gravity theory can give a comprehensive explanation of key cosmological phenomena, such as cosmic inflation and dark energy. During the early evolution of the Universe, modified gravity theories with geometry-matter coupling could have played a crucial role in the formation and evolution of large-scale structures. Generally, such theories can provide a unified framework for understanding the influence of gravity on the dynamics of the universe, not only during inflation but also during the late stages of the cosmological evolution.

In the present study, we have investigated in detail the cosmological implications of a specific model of $f(R,\mathcal{G},T)$ gravity, which is the linear combination of the GB invariant, of an arbitrary power of the Ricci scalar, and of the square root of the matter energy-momentum tensor. This specific choice of the Lagrangian density guarantees the covariant conservation of the matter energy-momentum tensor, which generally is not conserved in GB type gravities in the presence of geometry-matter coupling. The generalized Friedmann equations of this theory contain supplementary terms that can be interpreted as an effective dark energy of purely geometric origin. Even for a Universe filled with pressureless dust, the redshift evolution of the Hubble function, expressed as a second-order strongly nonlinear differential equation given by Eq.~(\ref{hubble_full}), is extremely complicated from a mathematical point of view, and, to investigate the properties of the model, the extensive use of numerical methods is necessary. After obtaining the solution of Eq.~(\ref{hubble_full}) numerically, a detailed comparison of the theoretical predictions of the model with observations was performed. Three distinct observational data sets were used for the comparison with the observations (Cosmic Chronometers, Type Ia supernovae, and Baryon Acoustic Oscillations data). A detailed comparison with the standard $\lambda$CDM paradigm was also considered,  with the statistical relevance of the results estimated using several statistical parameters (AIC, BIC, p). A general conclusion of this comparison is that the model fits the cosmological data well and reproduces almost perfectly the $\Lambda$CDM predictions at least in the redshift range $z\in [0,2.5]$. However, the $\Lambda$CDM model may be favored from an observational point of view due to the lower number of free parameters it has, but, on the other hand, it faces the very serious problem of the lack of a firm theoretical explanation of the nature of the cosmological constant.

One of the important mathematical properties of modified gravity theories is that their generalized Friedmann equations can be formulated as systems of autonomous differential equations, since their only dependence is on a well-defined time variable, the cosmological time, or an alternative variable $N=\ln a(t)$. The dynamical systems approach offers a reliable way to study the qualitative properties of the generalized Friedmann equations in the specific model of gravity considered $f(R,\mathcal{G},T)$. By introducing a set of model-independent variables $\left(x_1,x_2,...,x_7\right)$, and by using the linear Lyapunov stability analysis, we have studied the qualitative properties of the modified gravity model $f(R,\mathcal{G},T)$. The variables $x_i$, $i1,...,7$ are associated not only with the sources of matter, such as $x_4$ and $x_5$, but are related to the Lagrange density $f$ and its various derivatives.  Thus, the choice of variables and the complexity of the dynamical system are directly related to the background gravitational model and its complex physical and geometrical structure.  

Hence, the model can be formulated in terms of a seven-dimensional first-order system of differential equations, which also contains a constraint equation. However, the formulation of the dynamical system allows better insight into the general properties of the model than using other approaches. On the other hand, the complexity of the dynamical system is indicated by the existence of eight critical points, which also underlines the complexity of the cosmological dynamic of the model. 

The results of the analysis of dynamical systems for the $f(R,\mathcal{G},T)$ model show that at the background level the model passes all cosmological observational tests, predicting, among others, the correct time behavior of the matter density parameter and of the effective equation of state. Moreover, there exists an early-time radiation-dominated stable era, a matter saddle, and an accelerating de Sitter-type attractor. However, the qualitative analysis of the evolution equations reveals the possibility of the existence of a much more complicated and rich dynamics than the radiation-matter-acceleration scenario. In particular, Chaplygin gas, quintessence or phantom dark energy scenarios and evolution phases of the Universe are also possible, with the model predicting a much richer dynamical evolution than general relativity in the presence of a simple cosmological constant.    

Hence, the results obtained in the present study give more reason to continue to study this model in the future. In particular, it would be interesting to see how the presence of a pressure term, for example in the form of radiation,  would modify the dynamical characteristics of the evolution of the system. The study of inflation would also be of interest in the present model, since an equivalent model of accelerating expansion of the early Universe could be constructed in purely geometric terms, without the need of introducing scalar fields. Thus, a more general approach could give the possibility of a better theoretical determination of the effective equation of state of the dark energy, and of the deceleration parameter. Moreover, it would be possible to exclude gravitational models that cannot pass the observational tests and to obtain better constraints on the free parameters of the models.

\section*{Acknowledgments}
RM is thankful to UGC, Govt. of India, for providing Senior Research Fellowship (NTA Ref. No.: 211610083890).

\end{document}